\documentclass[aps,twocolumn,prb,showpacs,superscriptaddress]{revtex4-1}
\usepackage{dcolumn}
\usepackage{bm}

\newcommand{\figref}[1]{\figurename~\ref{#1}}
\newcommand*\diff{\mathop{}\!\mathrm{d}}

\usepackage{braket}

\usepackage[T1]{fontenc}
\usepackage[utf8]{inputenc} 
\usepackage{marvosym}
\usepackage{graphicx}
\usepackage{rotating}
\usepackage{amsmath}
\usepackage{amssymb}
\usepackage{booktabs}
\usepackage{url}

\usepackage{enumitem}
\setcitestyle{numbers,square}

\usepackage{setspace}
\usepackage{float} 

\usepackage{hyperref} 
\hypersetup{
    colorlinks=true,
    linkcolor=black,    
    citecolor=black,    
    urlcolor=blue,      
}

\begin{document}


\title{Superfluid Drag in Multicomponent Bose-Einstein Condensates on a Square Optical Lattice}

\author{Stian Hartman}
 \affiliation{Department of Physics, Norwegian University of Science and Technology, NO-7491 Trondheim, Norway}
 
\author{Eirik Erlandsen}
 \affiliation{Department of Physics, Norwegian University of Science and Technology, NO-7491 Trondheim, Norway}

\author{Asle Sudbø}
 \affiliation{Department of Physics, Norwegian University of Science and Technology, NO-7491 Trondheim, Norway}
 \affiliation{Center for Quantum Spintronics, Norwegian University of Science and Technology, NO-7491 Trondheim, Norway}
 
\date{\today}

\begin{abstract}
The superfluid drag-coefficient of a weakly interacting three-component Bose-Einstein condensate is computed on a square optical lattice deep in the superfluid phase, starting from a Bose-Hubbard model with component-conserving, on-site interactions and nearest-neighbor hopping. At the mean-field level, Rayleigh-Schr{\"o}dinger perturbation theory is employed to provide an analytic expression for the drag density. In addition, the Hamiltonian is diagonalized numerically to compute the drag within mean-field theory at both zero and finite temperatures to all orders in inter-component interactions. Moreover, path integral Monte Carlo simulations, providing results beyond mean-field theory, have been performed to support the mean-field results. In the two-component case the drag increases monotonically with the magnitude of the inter-component interaction $\gamma_{AB}$ between the two components $A$ and $B$. The increase is independent of the sign of the inter-component interaction.  This no longer holds when an additional third component $C$ is included. Instead of increasing monotonically, the drag can either be strengthened or weakened depending on the details of the interaction strengths, for weak and moderately strong interactions. The general picture is that the drag-coefficient between component $A$ and $B$ is a non-monotonic function of the inter-component interaction strength $\gamma_{AC}$ between $A$ and a third component $C$. For weak $\gamma_{AC}$ compared to the direct interaction $\gamma_{AB}$ between $A$ and $B$, the drag-coefficient between $A$ and $B$ can {\it decrease}, contrary to what one naively would expect. When $\gamma_{AC}$ is strong compared to $\gamma_{AB}$, the drag between $A$ and $B$ increases with increasing $\gamma_{AC}$, as one would naively expect. We attribute the subtle reduction of $\rho_{d,AB}$ with increasing $\gamma_{AC}$, which has no counterpart in the two-component case, to a renormalization of the inter-component scattering vertex $\gamma_{AB}$ via intermediate excited states of the third condensate $C$. We briefly comment on how this generalizes to systems with more than three components.     
\end{abstract}

\pacs{Valid PACS appear here}
\maketitle


\section{Introduction}
Following the experimental realization of Bose-Einstein condensation (BEC) in 1995 \cite{Anderson1995, Davis1995}, cold atomic gases have been intensely investigated both theoretically and experimentally. 
These systems are attractive to work with due to their high degree of tunability and absence of impurities, which are rare features of naturally occurring condensed matter systems. Magnetic and optical traps provide a high degree of control over the particles, and optical lattices with tunable depth and periodicity present an ideal arena for exploring a wide range of phenomena in condensed matter systems \cite{Bloch2005}. Especially intriguing is the possibility of introducing additional components in the system, leading to new dynamics and fascinating phenomena. Such multi-component systems can be produced in experiments by including different atoms, different isotopes of the same atom, or the same atom in different hyperfine states \cite{Multi1, Multi2, Multi3, Multi4}, and have been found to harbour rich physics. For a multi-component mixture of homo-nuclear atoms in different hyperfine states, the effects of strong spin-orbit coupling can be investigated by introducing a synthetic spin-orbit coupling by inducing transitions between the various hyperfine states \cite{Spielman2011}. A special case is when an entire hyperfine multiplet is present, making the internal degree of freedom an actual spin degree of freedom \cite{Multi4}. The focus of this paper will, however, be a general multi-component system in the absence of component-mixing interactions, as exemplified by the case of a heterogeneous mixture of different atoms.

In addition to the Mott insulating phases, BECs residing on an optical lattice can also exhibit superfluid properties. Superfluidity is in its own right a fascinating phenomenon, being a macroscopic manifestation of quantum mechanics, but there is a wide variety of physics present. Of particular interest is the multi-component case where interaction between particles of different types produces a dissipationless drag between the superfluid densities associated with each component, the so-called Andreev-Bashkin effect \cite{Andreev1975}. This drag affects the superfluid mass flow of the components, leading, for instance, to a superflow of one component inducing a superflow of a different component. The effect was initially studied in the context of a mixture of superfluid $\text{He}^4$ and $\text{He}^3$, but due to the low miscibility of such a mixture, this system has proven to be a poor candidate for investigating the drag \cite{Nespolo2017}. Presently, however, the sort of multi-component superfluids that are needed to produce the effect are realizable in cold atom systems and are in fact routinely made. The importance of the superfluid drag is rooted in the fact that superfluid systems depend strongly on the formation and interaction of quantum vortices \cite{Onsager1949, Feynman1955}, which are considerably influenced by drag interactions \cite{Dahl2008, Dahl2008v2, Dahl2008v3, Kaurov2005}. The presence of multiple components leads to a wider range of possible vortices, including drag-induced composite vortices, and significantly enriches the physics of topological phase transitions in superfluids as well as superconductors \cite{Dahl2008, Dahl2008v2, Dahl2008v3, Kaurov2005, Smiseth2005}.

For two-component Bose gases the microscopic origin of the drag interactions has been investigated in the weakly interacting limit, through mean-field theory, in both free space and on lattices \cite{Fil2005, Asle2009, Yanay2012, Hofer2012}. In addition, Quantum Monte Carlo simulations have been performed to explore the system in both the weak and strongly coupled regimes \cite{Kaurov2005, Lingua2015, Babaev2018}. The effect of a third component has, however, yet to be considered to any detailed degree. The objective of this paper is therefore to investigate such a three-component system and determine the dependence of the drag on the sign and strength of the inter-component interactions.

\section{Superfluid Drag Coefficients}\label{SF_drag}
In Landau's theory of superfluids there are two components associated with the fluid: One component moves independently of its container at equilibrium and experiences no loss of energy, and the second "normal" component moves with the container due to friction along the walls \cite{Landau1941}. The densities of the super and the normal components can be defined by considering a fluid inside an infinitely long cylinder, both of which are initially at rest. Providing the container with a small velocity -$\bm{v}_s$ and changing to the frame in which the cylinder is at rest (moving with velocity $\bm{v}_s$ relative to the initial frame) when equilibrium is reached will yield a superfluid mass current $\bm{j}_s$
\begin{equation}
    \bm{j}_s = \rho_s \bm{v}_s.
\end{equation}
where $\rho_s$ is the superfluid density  of the system, and $\bm{v}_s$ is the superfluid velocity. This equation describes a superfluid mass that moves independently of the container. The normal density $\rho_n$, which is at rest relative to the cylinder due to friction at the boundaries, is therefore defined as
\begin{equation}
    \rho_n = \rho - \rho_s,
\end{equation}
where $\rho$ is the total density of the fluid. The free energy density $\mathcal{F}$  in the frame moving with the cylinder is the free energy of the system at rest $\mathcal{F}_0$ plus the kinetic energy of the superfluid mass current,
\begin{equation}
    \mathcal{F} = \mathcal{F}_0 + \frac{1}{2}\rho_s \bm{v}^2_s.
\end{equation}
From $\mathcal{F}$ we may obtain the superfluid current density $\bm{j}_s$ as follows
\begin{equation}
    \bm{j}_s = \frac{\partial \mathcal{F}}{\partial \bm{v}_s}.
\end{equation}
The superfluid density $\rho_s$ is correspondingly obtained by,
\begin{equation}
    \rho_s = \frac{\partial^2 \mathcal{F}}{\partial \bm{v}^2_s}\Big|_{v_s\rightarrow 0}.
\end{equation}

In two-component systems, i.e.\ two types of bosons, there can be two superfluid densities and a non-dissipative drag between them. Such a drag originates with elastic momentum transfer from one species of atoms to another mediated by weak van der Waals forces \cite{Andreev1975}. The free energy density with very small superfluid velocities of two components $A$ and $B$ now reads \cite{Andreev1975}
\begin{equation}
    \begin{split}
        \mathcal{F} = & \mathcal{F}_0 + \frac{1}{2} \Big[(\rho_{nA} + \rho_{nB})\bm{v}^2_{n} + \rho_{sA}\bm{v}^2_{sA} \\
        & + \rho_{sB}\bm{v}^2_{sB} + \rho_d(\bm{v}_{sA} - \bm{v}_{sB})^2 \Big],
    \end{split}
\label{eq:superfluid_densities_from_F}
\end{equation}
where a finite normal velocity $\bm{v}_n$ with density $\rho_{nA} + \rho_{nB}$ is included. The notation is similar to the one used for the one-component case, now extended to the two components $A$ and $B$. The dissipationless drag between components $A$ and $B$ is quantified by the superfluid drag density $\rho_d$ and can be found by
\begin{equation}
    \rho_d = \frac{\partial^2 \mathcal{F}}{\partial \bm{v}_{sA}\partial \bm{v}_{sB}}\Big|_{v_{sA},v_{sB}\rightarrow 0}.
\end{equation}
In the two-component case, the superfluid mass current becomes
\begin{equation}
    \bm{j}_{sA} = \frac{\partial \mathcal{F}}{\partial \bm{v}_{sA}} = (\rho_{sA} - \rho_d)\bm{v}_{sA} + \rho_d\bm{v}_{sB},
\label{eq:superfluid_mass_current_with_drag}
\end{equation}
i.e.\ the superfluid mass current of one component can induce a co-directed ($\rho_d > 0$) or a counter-directed ($\rho_d < 0$) superfluid mass current of the other component.

The superfluid velocity is related to the phase of the superfluid order parameter, $\psi_{0\alpha}(\bm{r}) = \big<\hat{\psi}_{\alpha}(\bm{r})\big> = \psi_{0\alpha}e^{i\Theta_{\alpha}(\bm{r})}$ through $\bm{v}_{s\alpha}=\bm{\nabla}\Theta_{\alpha}/m_{\alpha}$, where $m_{\alpha}$ is the mass of component $\alpha$, and can be introduced by imposing twisted boundary conditions on the system \cite{Weichman1988}. 
Here, $\psi_{0\alpha}(\bm{r})$ is the local wavefunction of component $\alpha$ of the macroscopic condensate describing the multi-component superfluid. (We return to a more precise definition of this order parameter below). Here, $\alpha$ runs over the number of components of the system.  
The twisting of the phase of the superfluid order parameter is done by adding the factor $e^{i\bm{k}_{0\alpha}\cdot\bm{r}}$ to the field operator for component $\alpha$, so that $e^{-i\bm{k}_{0\alpha}\cdot\bm{r}}\hat{\psi}_{\alpha}(\bm{r})$ obeys the usual periodic boundary condition in all directions. Given this choice of phase twist, the superfluid velocity of particles with mass $m_{\alpha}$ (comprising component $\alpha$) is given by $\bm{v}_{s\alpha}=\bm{k}_{0\alpha}/m_{\alpha}$, and the expression for the drag density becomes
\begin{equation}
	\rho_{d} = m_{A} m_{B} \left( \frac{\partial^2 \mathcal{F}}{\partial \bm{k}_{0A}\partial \bm{k}_{0B}} \right)_{k_{0A}, k_{0B}\rightarrow 0}.
\end{equation}

In systems defined on a lattice rather than on a continuum, the drag density may depend on Cartesian direction, and $\rho_d$ will instead be a tensor of rank 2 \cite{Yanay2012},
\begin{equation}
	\rho^{ij}_d = m_{A} m_{B}\left( \frac{\partial^2 \mathcal{F}}{\partial k_{0A i}\partial k_{0B j}} \right)_{k_{0A}, k_{0B}\rightarrow 0},
\end{equation}
where $(i,j)$ are Cartesian lattice directions. On a square lattice, which is the case considered in this paper, we have $(i,j) \in(x,y)$. Furthermore, $k_{0\alpha i}$ is $i$-th Cartesian component of the twist-vector $\bm{k}_{0 \alpha}$. On a square lattice, we have $\rho^{ij}_d = \delta_{ij}\rho_d$ and the lattice vector can be aligned along the coordinate axes. The Cartesian direction $i$ is therefore chosen to be along the $x$-axis. Hence, the superfluid drag density on the square lattice may be obtained from the expression
\begin{equation}
	\rho_d = m_{A} m_{B} \left( \frac{\partial^2 \mathcal{F}}{\partial k_{0Ax}\partial k_{0Bx}} \right)_{k_{0A}, k_{0B}\rightarrow 0}.
\label{eq:Weichman_method_on_lattice}
\end{equation}

The generalization to three components $(A,B,C)$ is straightforward, and the free energy reads
\begin{equation}
    \begin{split}
	    \mathcal{F} = &\mathcal{F}_0 + \frac{1}{2} \big[(\rho_{nA} + \rho_{nB} + \rho_{nC})\bm{v}^2_{n} \\
	    &+ \rho_{sA}\bm{v}^2_{sB} + \rho_{sB}\bm{v}^2_{sB} + \rho_{sC}\bm{v}^2_{sC} \\
	    &- \rho_{d,AB}(\bm{v}_{sA} - \bm{v}_{sB})^2 \\
	    &- \rho_{d,AC}(\bm{v}_{sA} - \bm{v}_{sC})^2 \\
	    &- \rho_{d,BC}(\bm{v}_{sB} - \bm{v}_{sC})^2 \big].
	\end{split}
\label{eq:superfluid_densities_from_F_3_comp}
\end{equation}
There are now three superfluid drag densities quantifying the drag between each pair of boson components. The superfluid mass currents, densities, and drags can be found as before,
\begin{equation}
    \begin{split}
        \bm{j}_{sA} = \frac{\partial \mathcal{F}}{\partial \bm{v}_{sA}}
        = & (\rho_{sA} - \rho_{d,AB} - \rho_{d,AC})\bm{v}_{sA} \\
        & + \rho_{d,AB}\bm{v}_{sB} + \rho_{d,AC}\bm{v}_{sC},
    \end{split}
\label{eq:superfluid_flow_drag_density}
\end{equation}
\begin{equation}
	\rho_{sA} = \frac{\partial^2 \mathcal{F}}{\partial \bm{v}_{sA}^2}\big|_{v_{sA}\rightarrow 0},
\end{equation}
\begin{equation}
	\rho_{d,AB} = \frac{\partial^2 \mathcal{F}}{\partial\bm{v}_{sA}\partial \bm{v}_{sB}}\big|_{v_{sA},v_{sB}\rightarrow 0},
\end{equation}
and similarly for the remaining currents and densities.

\section{Mean-Field Model}\label{MF}
The starting point of our treatment is the $N$-component Bose-Hubbard model formulated in the grand canonical ensemble for a system with nearest-neighbor hopping and on-site interactions,
\begin{equation}
    H = H_1 + H_2.
    \label{B-H}
\end{equation}
Here, $H_1$ is the one-particle part of the Hamiltonian, 
\begin{equation}
        H_1 = -\sum_{\alpha} \Big\{  t_{\alpha}\sum_{\langle i,j \rangle} a^{\dagger}_{\alpha i} a_{\alpha i+j} 
         + \mu_{\alpha}\sum_i a^{\dagger}_{\alpha i} a_{\alpha i} \Big\},
         \label{B-H-1}
\end{equation}
while the interaction part of the Hamiltonian is given by 
\begin{equation}
    H_2 = \frac{1}{2} \sum_{\alpha \beta} \sum_{i} \gamma_{\alpha \beta} a^{\dagger}_{\alpha i} a^{\dagger}_{\beta i} a_{\beta i} a_{\alpha i}.
    \label{B-H-2}
\end{equation}
Here, $\sum_{\alpha}$ runs over the components of the system, $\sum_i$ runs over all $N_s$ lattice sites, $\sum_{\langle i,j \rangle}$ is a 
sum over lattice sites $i$ and their nearest neighbors $j$, $t_{\alpha}$ is the nearest-neighbor hopping matrix element on the square lattice 
for component $\alpha$, and $\mu_{\alpha}$ is a chemical potential determining the filling fraction of particles of component $\alpha$ on the lattice.
Moreover, $\gamma_{\alpha \beta}$ is the on-site (Hubbard) interaction between component $\alpha$ and $\beta$, which therefore includes intra-component ($\alpha=\beta $) as well as inter-component ($\alpha \neq \beta$) on-site interactions. Throughout, we consider repulsive on-site intra-component interactions $\gamma_{\alpha \alpha} > 0$, while the inter-component interactions can be both repulsive and attractive. We will define the intra-component Hubbard interactions $\gamma_{\alpha \alpha} 
\equiv \gamma_{\alpha} = \gamma$. In our numerical results to be presented below, we will furthermore set 
$t_{\alpha} = \gamma = 1$ as our units of energy.
The operators $a^{\dagger}_{\alpha i}$ and $a_{\alpha i}$ create and destroy bosons of component  $\alpha$ at lattice site $i$, respectively. The lattice constant of the square lattice, denoted by $d$, is generally taken to be unity. 

The BEC order parameter, $\psi_{0 \alpha i}$, and the operator describing fluctuations away from the condensate, $\hat{\phi}_{\alpha i}$, are introduced by defining
\begin{equation}
    \big< a_{\alpha i} \big> = \psi_{0 \alpha i}, \quad \hat{\phi}_{\alpha i} = a_{\alpha i} - \psi_{0 \alpha i}.
\end{equation}
The phase twist $\bm{k}_{0\alpha}$ is imposed on the order parameter, $\psi_{0 \alpha i} = \sqrt{n_{0\alpha}}e^{i\bm{k}_{0 \alpha} \cdot \bm{r}_i}$, where $n_{0\alpha}$ is identified as the condensate density of component $\alpha$. The order parameter and the new lattice operator are inserted into the Hamiltonian and terms more than quadratic in $\hat{\phi}_{\alpha i}$ neglected. The new operator is Fourier transformed with the twisted boundary conditions,
\begin{equation}
    \hat{\phi}_{\alpha i} = \frac{1}{\sqrt{N_s}} \sum_{\bm{k}\neq 0} e^{i(\bm{k} + \bm{k}_{0 \alpha}) \cdot \bm{r}_i} b_{\bm{k} \alpha},
\end{equation}
using that $\bm{r}_{i \pm \eta} = \bm{r}_i \pm \bm{\eta}$, and the relation
\begin{equation}
    \frac{1}{N_s} \sum_{i} e^{i\bm{k}\cdot\bm{r}_i} = \delta_{\bm{k}, 0}.
\end{equation}
Here, $\bm{\eta}$ is a unit vector connecting nearest neighbors on the lattice, and $\delta_{\bm{k},0}$ is the Kronecker delta, which is unity if $\bm{k} = 0$, and zero otherwise. 
For $\hat{\phi}_{\alpha i}$ to have twisted boundary conditions the momentum $\bm{k}$ must obey periodic boundary conditions.
The mean-field Hamiltonian then takes the form
\begin{equation}
    H = \widetilde{H}_0 + \widetilde{H}_2,
\end{equation}
where the first part contains the zeroth-order terms in the operators,
 \begin{equation}
    \begin{split}
        \widetilde{H}_0 = N_s\Big\{ &-\sum_{\alpha} n_{0\alpha}^2 \Big[ \mu_{\alpha} + 2t_\alpha\sum_{\bm{\eta}}\cos\big(\bm{k}_{0\alpha}\cdot\bm{\eta}\big) \Big] \\
        & + \frac{1}{2}\sum_{\alpha\beta}\gamma_{\alpha\beta}n^2_{0\alpha} n^2_{0\beta} \Big\},
    \end{split}
 \label{eq:mean_field_H0}
 \end{equation}
 and the second bilinear term takes the form
 \begin{equation}
    \begin{split}
        \widetilde{H}_2 = & \sum_{\bm{k}\neq 0} \Big\{ -\sum_{\alpha} \Big[ \mu_{\alpha} + 2t_\alpha\sum_{\bm{\eta}}\cos\big((\bm{k}+\bm{k}_{0\alpha})\cdot\bm{\eta}\big) \Big] b^{\dagger}_{\bm{k}\alpha}b_{\bm{k}\alpha} \\ & + \sum_{\alpha\beta}\frac{\gamma_{\alpha\beta}}{2} \Big[ \sqrt{n_{0\alpha} n_{0\beta}} \left( 2 b^{\dagger}_{\bm{k}\beta}b_{\bm{k}\alpha} + b_{\bm{k}\alpha}b_{\bm{-k}\beta} + b^{\dagger}_{\bm{k}\alpha}b^{\dagger}_{\bm{-k}\beta} \right)\\
        &\quad\quad+ 2n^2_{0\alpha}b^{\dagger}_{\bm{k}\beta}b_{\bm{k}\beta} \Big] \Big\}.
    \end{split}
\label{eq:mean_field_H2}
\end{equation}
In this form the excitation spectrum will contain the chemical potential, and the condensate densities $n_{0\alpha}$ should be determined self-consistently by minimizing the free energy when $\mu_{\alpha}$ is given. However, $\mu_{\alpha}$ can be removed from the bilinear terms by using that the total number density of boson type $\alpha$ is
\begin{equation}
    n_{\alpha} = n_{0\alpha} + \frac{1}{N_s}\sum_{\bm{k}\neq 0}b^{\dagger}_{\bm{k}\alpha}b_{\bm{k}\alpha}.
\label{eq:total_numer_density_relation}
\end{equation}
Inserting this into $\widetilde{H}_0$, once more neglecting terms that are more than quadratic in operators, gives the substitution $n_{0\alpha}\rightarrow n_{\alpha}$ in addition to extra bilinear terms. Using \eqref{eq:total_numer_density_relation} on $\widetilde{H}_2$ results in $n_{0\alpha}\rightarrow n_{\alpha}$. The two parts of the Hamiltonian are recast so that $\widetilde{H}_0$ again only contains terms of zeroth order in operators, and $\widetilde{H}_2$ is bilinear;
 \begin{equation}
    \begin{split}
        \widetilde{H}_0 = N_s\Big\{ &-\sum_{\alpha} n_{\alpha}^2 \Big[ \mu_{\alpha} + 2t_\alpha\sum_{\bm{\eta}}\cos\big(\bm{k}_{\alpha}\cdot\bm{\eta}\big) \Big] \\
        & + \frac{1}{2}\sum_{\alpha\beta}\gamma_{\alpha\beta}n^2_{\alpha} n^2_{\beta} \Big\},
    \end{split}
 \label{eq:mean_field_H0_new}
 \end{equation}
 
\begin{equation}
    \widetilde{H}_2 = \sum_{\alpha}\widetilde{H}_{\alpha} + \sum_{\alpha\neq\beta}\widetilde{H}_{\alpha\beta},
 \label{eq:mean_field_H2_new}
\end{equation}
\begin{equation}
    \begin{split}
        \widetilde{H}_{\alpha} = \sum_{\bm{k}\neq 0}&\Big\{ \left(E_{\bm{k}\alpha} + f_{\bm{k}\alpha}\right)b^{\dagger}_{\bm{k}\alpha}b_{\bm{k}\alpha}\\
        & + \frac{1}{2}F_{\alpha}\left( b_{\bm{k}\alpha}b_{-\bm{k}\alpha} + b^{\dagger}_{\bm{k}\alpha}b^{\dagger}_{-\bm{k}\alpha} \right) \Big\}
    \end{split}
\label{eq:single_component_terms_H2_new}
\end{equation}
\begin{equation}
    \widetilde{H}_{\alpha\beta} = \frac{1}{2}U_{\alpha\beta}\sum_{\bm{k}\neq 0}\Big\{2b^{\dagger}_{\bm{k}\alpha}b_{\bm{k}\beta} + b_{\bm{k}\alpha}b_{-\bm{k}\beta} + b^{\dagger}_{\bm{k}\alpha}b^{\dagger}_{-\bm{k}\beta} \Big\},
\label{eq:inter_component_terms_H2_new}
\end{equation}
where the coefficients are
\begin{equation}
    \begin{split}
        & E_{\bm{k}\alpha} = \epsilon_{\bm{k}\alpha} +  F_\alpha \\
        & \epsilon_{\bm{k}\alpha} = 2t_\alpha\sum_{\bm{\eta}}\left[1-\cos(\bm{k}\cdot\bm{\eta})\right]\cos(\bm{k}_{0\alpha}\cdot\bm{\eta}) \\
        & f_{\bm{k}\alpha} = 2t_\alpha\sum_{\bm{\eta}}\sin(\bm{k}\cdot\bm{\eta})\sin(\bm{k}_{0\alpha}\cdot\bm{\eta}) \\
    	& F_{\alpha} = \gamma_{\alpha}n_{\alpha} \\
        & U_{\alpha\beta} = \gamma_{\alpha\beta}\sqrt{n_{\alpha}n_{\beta}}.
    \end{split}
\label{eq:definition_H2_new_coefficients}
\end{equation}

The equations \eqref{eq:mean_field_H0_new}-\eqref{eq:definition_H2_new_coefficients} constitute the mean-field model for the weakly interacting Bose gas with $N$ components and twisted boundary conditions. In the following, the superfluid drag density is re-derived in the two-component case, and then the effect of a third component on the drag considered.

In the coherent state basis the path integral formulation of the partition function reads \cite{Negele1988}
 \begin{equation}
     \mathcal{Z} = \int\mathcal{D}\left[\left\{ \phi^*_{\Omega}(\tau), \phi_{\Omega}(\tau) \right\}\right] e^{-\int^{\beta}_{0}\diff\tau S},
\label{eq:path_integral_formulation_partition_function}
 \end{equation}
 where
 \begin{equation}
     S = \sum_{\Omega}\left[ \phi^*_{\Omega}\partial_{\tau}\phi_{\Omega} + H\left(\left\{ \phi^*_{\Omega}, \phi_{\Omega} \right\}\right) \right].
 \end{equation}
 
 In $H\left(\left\{ \phi^*_{\Omega}, \phi_{\Omega} \right\}\right)$ the Hamiltonian is normal ordered and the operators replaced by complex valued fields by the prescription $b_{\Omega} \rightarrow \phi_{\Omega}$  and $b^{\dagger}_{\Omega} \rightarrow \phi^*_{\Omega}$, with $\Omega$ labeling the quantum numbers, $\left\{\Omega\right\} = \left\{\bm{k},\alpha\right\}$. To satisfy the bose statistics the fields are periodic in imaginary time, $\phi_{\Omega}(\tau) = \phi_{\Omega}(\tau+\beta)$, where $\beta = 1/T$ is the inverse temperature.

\section{Drag in Two-Component BEC}\label{2-Comp}
In the two-component case the boson components present are $\alpha,\beta = A,B$. The energy spectrum, and hence the free energy, is known from the literature when the twist is zero \cite{Asle2009}, but the additional terms $f_{\bm{k}\alpha}$ yield large and unwieldy eigenenergies when diagonaliziing the Hamiltonian. Fortunately, since  $f_{\bm{k}\alpha}\rightarrow 0$ as $\bm{k}_{0\alpha}\rightarrow 0$, the path integral formulation of the partition function can be employed and expanded in powers of $f_{\bm{k}\alpha}$.

 Computing the path integral and expanding to second order in $f_{\bm{k}\alpha}$, which is done in appendix \ref{app:Computation_of_Path_Integral}, yields
 \begin{equation}
     \mathcal{Z} = \mathcal{Z}_1 \mathcal{Z}_2 = e^{-\beta \mathcal{F}}.
      \label{partition_function}
 \end{equation}
 Here, $\mathcal{Z}_1$ is the partition function due to the Hamiltonian when $f_{\bm{k}\alpha}=0$, which can be found by a general Bogoliubov tranformation to the diagonal basis $c_{\bm{k}\pm}$ (see appendix \ref{app:Diagonalization});
 
  \begin{equation}
    \begin{split}
        \mathcal{Z}_1 = e^{-\beta\widetilde{H}_0} & \prod_{\bm{k}\neq 0}  e^{-\frac{\beta}{2}(\mathcal{E}_{\bm{k}+} + \mathcal{E}_{\bm{k}-} - E_{\bm{k}A} - E_{\bm{k}B})} \\
        &\times \Big[1-e^{-\beta\mathcal{E}_{\bm{k}+}}\big]^{-1}\big[1-e^{-\beta\mathcal{E}_{\bm{k}-}}\big]^{-1}.
    \end{split}
 \end{equation}
 In \eqref{partition_function} $\mathcal{F}$ is the Helmholtz free energy. The energy spectrum is
 \begin{equation}
    \begin{split}
        \mathcal{E}_{\bm{k}\pm} & = \frac{1}{\sqrt{2}}\Big\{ \epsilon_{\bm{k}A}\left( \epsilon_{\bm{k}A} + 2 F_{A} \right) 
        + \epsilon_{\bm{k}B}\left( \epsilon_{\bm{k}B} + 2 F_{B} \right)\\
        & \quad \pm\Big\{ \big[\epsilon_{\bm{k}A}\left( \epsilon_{\bm{k}A} + 2 F_{A} \right) 
        - \epsilon_{\bm{k}B}\left( \epsilon_{\bm{k}B} + 2 F_{B} \right)\big]^2 \\
        &\quad\quad\quad + 16U_{AB}^2\epsilon_{\bm{k}A}\epsilon_{\bm{k}B} \Big\}^{\frac{1}{2}}\Big\}^{\frac{1}{2}}.
    \end{split}
\label{eq:two_component_eigenenergy}
\end{equation}
 
The expansion in powers of $f_{\bm{k}\alpha}$ is contained in $\mathcal{Z}_2$. Terms which are first order in $f_{\bm{k}\alpha}$ vanish, and only the term proportional to $f_{\bm{k}A}f_{\bm{k}B}$ in second order is non-zero during the differentiation in \eqref{eq:Weichman_method_on_lattice}. Only this term is therefore of interest with regards to the drag.
The free energy density is obtained by the usual relation
\begin{equation}
    \mathcal{F} = -\frac{1}{\beta N_s}\ln \mathcal{Z} = -\frac{1}{\beta N_s}\big(\ln \mathcal{Z}_1 + \ln \mathcal{Z}_2 \big),
\end{equation}
and is at $T=0$
\begin{equation}
	\begin{split}
		\mathcal{F} = & \frac{\widetilde{H}_0}{N_s} + \frac{1}{2N_s} \sum_{\bm{k}\neq 0}(\mathcal{E}_{\bm{k}+} + \mathcal{E}_{\bm{k}-} - E_{\bm{k}A} - E_{\bm{k}B}) \\
		&
		+ \frac{2 U^2_{AB}}{N_s}\sum_{\bm{k}\neq 0} f_{\bm{k}A}f_{\bm{k}B} \frac{\epsilon_{\bm{k}A} \epsilon_{\bm{k}B}}{\mathcal{E}_{\bm{k}+}\mathcal{E}_{\bm{k}-}(\mathcal{E}_{\bm{k}+} + \mathcal{E}_{\bm{k}-})^3} \\
		& + \mathcal{O}(f^2_{\bm{k}\alpha}),
	\end{split}
\end{equation}
which finally yields the superfluid drag density in the two-component case at zero temperature as
\begin{equation}
	\rho_d = \frac{8 m_A m_B t_A t_B d^2}{N_s}\sum_{\bm{k}\neq 0} 
	\frac{ U^2_{AB}\epsilon_{\bm{k}A} \epsilon_{\bm{k}B}\sin^2(k_x d)}{\mathcal{E}_{\bm{k}+}\mathcal{E}_{\bm{k}-}(\mathcal{E}_{\bm{k}+} + \mathcal{E}_{\bm{k}-})^3}.
\label{eq:superfluid_drag_density}
\end{equation}
This result agrees with previously published results obtained by slightly different methods \cite{Asle2009, Yanay2012}. A few details are worth noting. 
    The inter-component interaction strength $\gamma_{AB}$ only appears as $\gamma^2_{AB}$, so its sign does not matter. Both attractive and repulsive interactions between the two boson components yield the same positive superfluid drag density, meaning that the superfluid flow of one component induces a co-directed flow in the other, as seen from \eqref{eq:superfluid_mass_current_with_drag}.
    The energy spectrum \eqref{eq:two_component_eigenenergy} can become imaginary in some parameter regions, which implies an instability of the system. The requirement that the two boson components can coexist in the BEC can be shown to be 
    \begin{equation}
    \gamma_{A}\gamma_{B} > \gamma^2_{AB}
    \label{stability-two-components}
    \end{equation}
    by demanding that the energy spectrum is real when $n_{\alpha} > 0$, or more precisely, when the expression inside the outer root of \eqref{eq:two_component_eigenenergy} is positive. The same criterion is obtained by minimizing $\widetilde{H}_0$ with respect to $n_{A}$ and $n_{B}$ and demanding $n_{\alpha}>0$.

\section{Drag in Three-Component BEC}\label{3-Comp}
In the three-component case, we have $\alpha,\beta = A, B, C$. For this case, an exact analytic expression analogous to \eqref{eq:superfluid_drag_density}, for the superfluid drag density cannot be obtained even at the mean-field level. However, the effect of a third component on the drag can be investigated by numerical diagonalization to all orders in the interaction strengths, as well analytically using inter-component interactions as perturbation parameters. 

In this section, we first diagonalize the Hamiltonian defined by 
\eqref{eq:mean_field_H2_new},
\eqref{eq:single_component_terms_H2_new}, and
\eqref{eq:inter_component_terms_H2_new} numerically to all orders in inter-component interactions, from which the free energy and drag density may be computed numerically. As a check on these results, we reproduce the analytical results for the two-component case when two of the three interaction parameters are set equal to zero.   
Secondly, we  use Rayleigh-Schrödinger perturbation theory up to fourth order in the inter-component interaction terms \eqref{eq:inter_component_terms_H2_new} to find their contribution to the free energy, from which the drag is obtained by \eqref{eq:Weichman_method_on_lattice}, resulting in an analytic perturbative expression for the drag. 
The former approach has the advantage of giving exact results within the weakly interacting mean-field model at both zero and finite temperatures. The latter has the advantage of providing an analytic expression that can be inspected to find the qualitative behaviour of the drag between $A$ and $B$ due to $C$ at zero temperature $T=0$. The accuracy of the perturbation approach may be checked by comparing with the numerical mean-field results, and we will generally find a surprisingly good agreement.  

Finally, in the third part of this section, we compute the drag-coefficients using path integral Quantum Monte Carlo simulations to compute the drag beyond mean-field theory, to substantiate the results found at the mean-field level. The results in this third part of the section are obtained at a low temperature $T=0.1$. 

\subsection{Weak-Coupling Mean-Field Theory: Numerical Diagonalization to All Orders in Inter-Component Interactions}\label{SubsectionA}
In the numerical approach the energy spectrum of \eqref{eq:mean_field_H2_new} is found numerically with $f_{\bm{k}\alpha}\neq 0$. The details of this computation are relegated to appendix \ref{app:Diagonalization}. From the eigenvalues of \eqref{eq:mean_field_H2_new},  the free energy is obtained by the  relation \eqref{partition_function}. The superfluid drag is then computed by a finite difference approximation of \eqref{eq:Weichman_method_on_lattice}. In the two-component case this approach reproduces the result of \eqref{eq:superfluid_drag_density} exactly.

\begin{figure*}[!ht]
 	\centering
 	\includegraphics[scale=0.48]{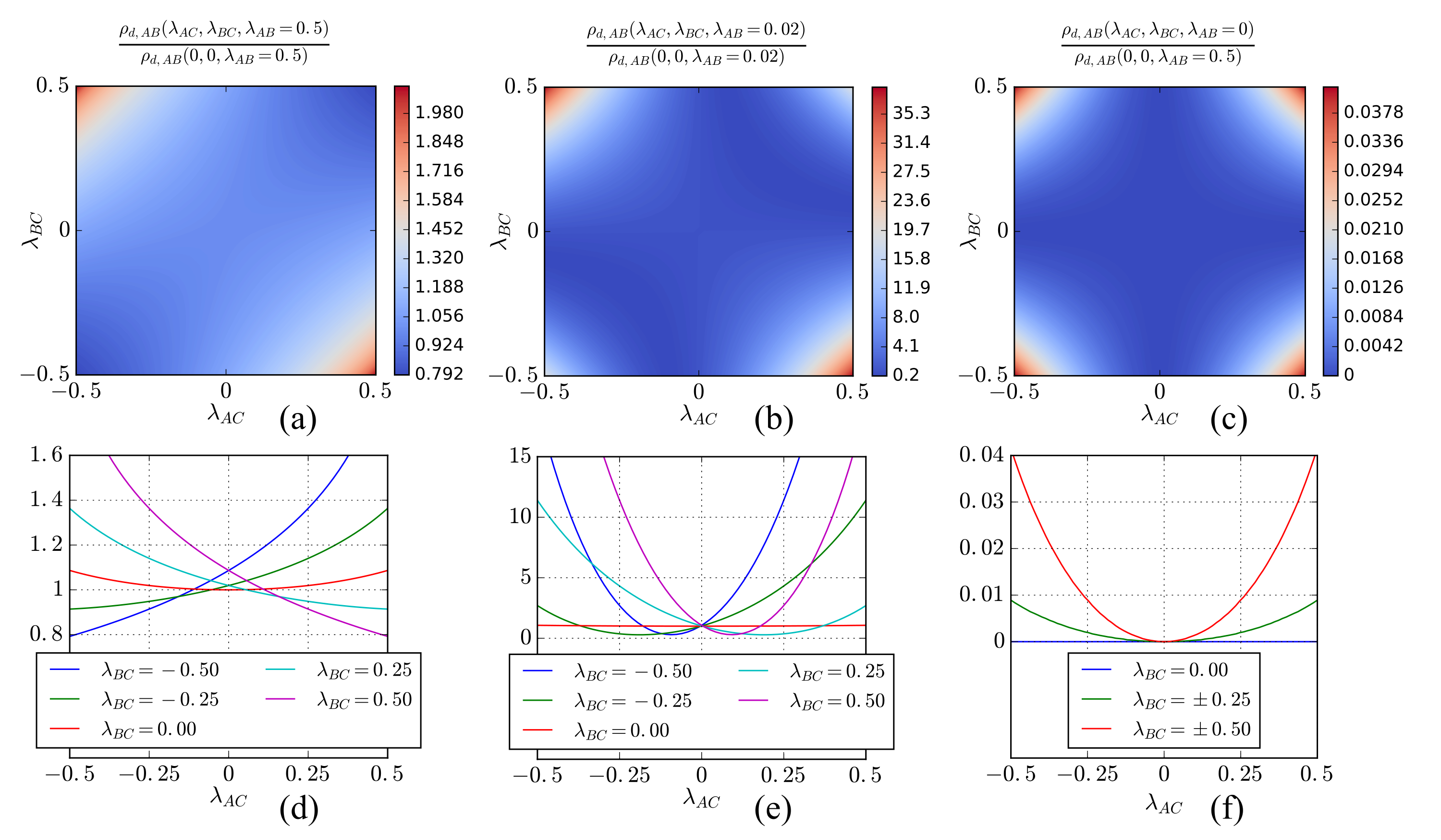}
 	\caption{The effect of a third component $C$ on the drag $\rho_{d,AB}$ from the mean-field numerical diagonalization at zero temperature, $T=0$, is displayed. The parameters are $N_s=10^4$, $d=1$, $t_{\alpha} = 1$, $m_{\alpha} = 1$, $\gamma_{\alpha} = \gamma = 1$, $n_{\alpha}=0.3$ for $\alpha=A,B,C$, where we have defined $\lambda_{\alpha\beta}=\gamma_{\alpha\beta}/\gamma$. In all cases, we have $\rho_{d,AB} \geq 0$. This is a generic feature of the weak-coupling limit. }
 	\label{fig:Three_Component_Drag_Numerical}
 \end{figure*}

The result of the numerical approach at zero temperature, which is expected to be quantitatively correct within the weakly interacting superfluid regime in which mean-field theory is valid, is presented in \figref{fig:Three_Component_Drag_Numerical}. 
 
The main feature to note is the {\it decreasing or non-monotonic} behavior of the inter-component drag between components $A$ and $B$, $\rho_{d,AB}$,  as a function of the inter-component interaction strength between $A$ and $C$, $\lambda_{AC}$. Here, we have defined $\lambda_{\alpha\beta}=\gamma_{\alpha\beta}/\gamma$. As the interaction strength $\lambda_{AC}$ {\it increases} in absolute value, there exists a parameter regime where the inter-component drag  $\rho_{d,AB}$ {\it decreases}. Such a result has no counterpart in the two-component case where, as we have emphasized and as was seen from  \eqref{eq:superfluid_drag_density}, the inter-component drag increases monotonically with inter-component interaction strength. Therefore, we conclude that the inter-component drag between components $A$ and $B$ mediated by an elastic momentum transfer via a third component $C$ has qualitatively new features. Eq. \eqref{eq:superfluid_drag_density} has one notable property, namely that the strength of the inter-component interaction enters as an even power (squared), which renders the corresponding drag always positive and monotonically increasing as a function of $\lambda_{AB}$. Such a direct contribution to $\rho_{d,AB}$ from $\lambda_{AB}$ clearly also exists in the present case, but there will be corrections to this as interactions via a third component are brought in. 

Since three interaction strengths now are involved, there could be corrections to the direct interaction $\lambda_{AB}$ giving the main contribution to $\rho_{d,AB}$ which would involve higher order terms in the interactions strengths, possibly of odd order. 
Although the direct term proportional to $\lambda_{AB}^2$ yields an exclusively a positive contribution to   $\rho_{d,AB}$ in the weak-coupling case considered in this paper, the corrections terms stemming from indirect terms via a third component, need not necessarily be positive. 

This crudely provides an insight into the results shown in  \figref{fig:Three_Component_Drag_Numerical} (a,d), (b,e), and (c,f). In \figref{fig:Three_Component_Drag_Numerical} (a,d), where the direct interaction $\lambda_{AB} = 0.50$, the qualitatively new feature compared to the two-component case is the {\it decrease} of $\rho_{d,AB}$ as a function of $\lambda_{AC}$ for $\lambda_{BC} = 0.25$ and $0.50$, as well as the decrease of $\rho_{d,AB}$ as  $|\lambda_{AC}|$ increases, for $\lambda_{BC} = -0.25$ and $-0.50$. The results for $\rho_{d,AB}$ are even in $\lambda_{AC}$ provided that we also switch the sign of $\lambda_{BC}$ when $\lambda_{AC}$ changes sign.

In \figref{fig:Three_Component_Drag_Numerical} (b,e), where the direct interaction is much weaker than in (d), $\lambda_{AB} = 0.02$, the same initial decrease as noted above persists, but now only up to much smaller values of $\lambda_{AC}$ before $\rho_{d,AB}$ eventually starts exhibiting the expected monotonic increase with increasing absolute value of $\lambda_{AC}$. The symmetry noted above under the simultaneous interchange of the signs of $\lambda_{AC}$ and $\lambda_{BC}$ also persists. 

In \figref{fig:Three_Component_Drag_Numerical} (c,f), where  $\lambda_{AB} = 0.00$, there is no longer any tendency to initial decrease as a function of $\lambda_{AC}$ for any value of $\lambda_{AB}$. It is also clear from these results that the value of $\lambda_{AB}$ sets a scale for the maximum values of $\lambda_{AC}$ up to which one can see a decrease in $\rho_{d,AB}$ as a function of increasing absolute value of $\lambda_{AC}$.   
These findings warrant a more detailed investigation. In order to gain further insight into what is going on in this case, we proceed by analyzing the corrections to the drag analytically in perturbation theory.  

\subsection{Weak-Coupling Mean-Field Theory: Perturbation Theory}\label{SubsectionB}
In the Rayleigh-Schrödinger perturbation scheme the Hamiltonian is separated as in \eqref{eq:mean_field_H2_new}, where the single-component part is diagonalized exactly by performing the usual Bogoliubov transformation
\begin{equation}
    b_{\bm{k}\alpha} = u_{\bm{k}\alpha}c_{\bm{k}\alpha} - v_{\bm{k}\alpha}c^{\dagger}_{-\bm{k}\alpha},
\label{eq:bogoliubov_transformation}
\end{equation}
on each component separately, where $c_{\bm{k}\alpha}$ is the operator in the diagonal basis for the single-component part, and $u_{\bm{k}\alpha}$ and $v_{\bm{k}\alpha}$ are the transformation coefficients. Using the condition that the new operators must preserve the boson commutation relation,
\begin{equation}
    u_{\bm{k}\alpha}^2 - v_{\bm{k}\alpha}^2 = 1,
\label{eq:bogoliubov_commutation_condition}
\end{equation}
results in the diagonal Hamiltonian for the single-component terms
\begin{equation}
    H_{\text{sol}} = \sum_{\bm{k}\neq 0}\sum_{\alpha}\widetilde{\mathcal{E}}_{\bm{k}\alpha}c^{\dagger}_{\bm{k}\alpha}c_{\bm{k}\alpha},
\end{equation}
where
\begin{equation}
    \widetilde{\mathcal{E}}_{\bm{k}\alpha} = \sqrt{\epsilon_{\bm{k}\alpha}\left(\epsilon_{\bm{k}\alpha}+2 F_\alpha\right)} + f_{\bm{k}\alpha} = \widetilde{E}_{\bm{k}\alpha} + f_{\bm{k}\alpha},
\label{eq_Hsol_energies}
\end{equation}
and 
\begin{equation}
    u_{\bm{k}\alpha}^2 = \frac{1}{2}\left(1 + \frac{E_{\bm{k}\alpha}}{\widetilde{E}_{\bm{k}\alpha}}  \right), \quad 
    v_{\bm{k}\alpha}^2 = \frac{1}{2}\left(1- \frac{E_{\bm{k}\alpha}}{\widetilde{E}_{\bm{k}\alpha}}  \right).
\label{eq:bogoliubov_tranformation_coefficients}
\end{equation}
The term $f_{\bm{k}\alpha}$, which is odd in $\bm{k}$, is the only term that does not appear in the usual Bogoliubov transformation. It can be shown to transform as
\begin{equation}
    \sum_{\bm{k}\neq 0}f_{\bm{k}\alpha}b^{\dagger}_{\bm{k}\alpha}b_{\bm{k}\alpha} = \sum_{\bm{k}\neq 0}f_{\bm{k}\alpha}c^{\dagger}_{\bm{k}\alpha}c_{\bm{k}\alpha},
\end{equation}
since the transformation coefficients $u_{\bm{k}\alpha}$ and $v_{\bm{k}\alpha}$ are even in $\bm{k}$ and fulfill Eq. \eqref{eq:bogoliubov_commutation_condition}.

In the new basis, the inter-component interaction terms, i.e.\ the perturbation, is given by
\begin{equation}
    \begin{split}
        H_{\text{pert}} &= \sum_{\{\alpha\beta \} } \left[\widetilde{H}_{\alpha\beta} + \widetilde{H}_{\beta\alpha}\right] \\
        & = \sum_{\{\alpha\beta \}}\sum_{\bm{k}\neq 0}\widetilde{U}_{\bm{k},\alpha\beta}\big[ c^{\dagger}_{\bm{k}\alpha}c_{\bm{k}\beta} + c^{\dagger}_{\bm{k}\beta}c_{\bm{k}\alpha} \\
        &\quad\quad\quad\quad + c_{\bm{k}\alpha}c_{-\bm{k}\beta} + c^{\dagger}_{\bm{k}\alpha}c^{\dagger}_{-\bm{k}\beta} \big],
    \end{split}
\label{eq:Hpert_in_diagonal_basis}
\end{equation}
where
\begin{equation}
    \begin{split}
        \widetilde{U}_{\bm{k}\alpha\beta} &= U_{\alpha\beta}\big[ u_{\bm{k}\alpha}u_{\bm{k}\beta} + v_{\bm{k}\alpha}v_{\bm{k}\beta} - u_{\bm{k}\alpha}v_{\bm{k}\beta} - u_{\bm{k}\beta}v_{\bm{k}\alpha} \big] \\
        &=U_{\alpha\beta}\sqrt{\frac{\epsilon_{\bm{k}\alpha}\epsilon_{\bm{k}}}{\widetilde{E}_{\bm{k}\alpha}\widetilde{E}_{\bm{k}\beta}}}.
    \end{split}
\end{equation}
The notation $\{ \alpha\beta \}$ indicates that the sum is taken over all pairs of $\alpha$ and $\beta$, disregarding the order.

A criterion for the perturbation expansion to be useful is for the correction terms to be small compared to the energy levels,
\begin{equation}
    |\widetilde{U}_{\bm{k}\alpha\beta}| \ll |\widetilde{\mathcal{E}}_{\bm{k}\alpha} + \widetilde{\mathcal{E}}_{\bm{k}\beta}|.
\end{equation}
This yields in the limit of small $\bm{k}_{0\alpha}$ and $\bm{k}\rightarrow 0$ where the perturbation is expected to be worst,
\begin{equation}
    |\gamma_{\alpha\beta}|\Big(\frac{n_{\alpha}t_{\alpha}n_{\beta}t_{\beta}}{\gamma_{\alpha}\gamma_{\beta}}\Big)^{\frac{1}{4}} \ll \sqrt{\gamma_{\alpha}n_{\alpha}t_{\alpha}} + \sqrt{\gamma_{\beta}n_{\beta}t_{\beta}},
\end{equation}
which should hold for all pairs of the boson components present.

Computing the perturbations up to fourth order, given in appendix \ref{app:RS_perturbation_theory}, and differentiating according to \eqref{eq:Weichman_method_on_lattice} gives the perturbative expression for the superfluid drag density between boson components $A$ and $B$

\begin{widetext}

 \begin{equation}
     \rho_{d,AB} = \rho^{\text{(2)}}_{d,AB} + \rho^{\text{(3)}}_{d,AB} + \rho^{\text{(4)}}_{d,AB},
\label{eq:drag_density_perturbation_full}
 \end{equation}
 where

\begin{equation}
    \begin{split}
        \rho^{(2)}_{d,AB} = \frac{8 m_{A}m_{B}t_{A}t_{B}d^2}{N_s} \sum_{\bm{k}\neq 0}\sin^2(k_x d) \frac{\widetilde{U}^2_{\bm{k}AB}}{ (\widetilde{E}_{\bm{k}A} + \widetilde{E}_{\bm{k}B})^3},
    \end{split}
\label{eq:drag_second_order}
\end{equation}

\begin{equation}
    \begin{split}
        \rho^{(3)}_{d,AB} =& -\frac{8 m_{A}m_{B}t_{A}t_{B}d^2}{N_s}\sum_{\bm{k}\neq 0}\sin^2(k_x d)\widetilde{U}_{\bm{k}AB}\widetilde{U}_{\bm{k}AC}\widetilde{U}_{\bm{k}BC}
        \Bigg\{ \frac{2}{(\widetilde{E}_{\bm{k}A}+\widetilde{E}_{\bm{k}B})^3}\left[ \frac{1}{\widetilde{E}_{\bm{k}A}
        + \widetilde{E}_{\bm{k}C}} + \frac{1}{\widetilde{E}_{\bm{k}B} + \widetilde{E}_{\bm{k}C}} \right]\\
        & + \frac{1}{(\widetilde{E}_{\bm{k}A}+\widetilde{E}_{\bm{k}B})^2}\left[ \frac{1}{(\widetilde{E}_{\bm{k}A} + \widetilde{E}_{\bm{k}C})^2} + \frac{1}{(\widetilde{E}_{\bm{k}B} + \widetilde{E}_{\bm{k}C})^2} \right] 
         - \frac{1}{(\widetilde{E}_{\bm{k}A} + \widetilde{E}_{\bm{k}C})^2(\widetilde{E}_{\bm{k}B} + \widetilde{E}_{\bm{k}C})^2}\Bigg\},
    \end{split}
\label{eq:drag_third_order}
\end{equation}

\begin{equation}
    \begin{split}
        \rho^{\text{(4)}}_{d,AB} = & \frac{16t_{A}t_{B}m_{A}m_{B}d^2}{N_s}\sum_{\bm{k}\neq 0}\frac{\widetilde{U}^4_{\bm{k}AB}\sin^2(k_x d)}{(\widetilde{E}_{\bm{k}A} + \widetilde{E}_{\bm{k}B})^3}\Bigg[\frac{1}{\widetilde{E}_{\bm{k}A}\widetilde{E}_{\bm{k}B}} + \frac{3}{(\widetilde{E}_{\bm{k}A} + \widetilde{E}_{\bm{k}B})^2} \Bigg] \\
        &+ \frac{8t_{A}t_{B}m_{A}m_{B}d^2}{N_s}\sum_{\bm{k}\neq 0}\sin^2(k_x d)\Bigg\{
        \widetilde{U}^2_{\bm{k}AB}\big[\widetilde{U}^2_{\bm{k}AC} + \widetilde{U}^2_{\bm{k}BC}\big]
        \Bigg( \frac{3}{(\widetilde{E}_{\bm{k}A} + \widetilde{E}_{\bm{k}B})^4}\Bigg[ \frac{1}{\widetilde{E}_{\bm{k}A} + \widetilde{E}_{\bm{k}C}} + \frac{1}{\widetilde{E}_{\bm{k}B} + \widetilde{E}_{\bm{k}C}} \Bigg] \\
        & + 
        \Bigg[\frac{1}{(\widetilde{E}_{\bm{k}A} + \widetilde{E}_{\bm{k}B})^3} - 
        \frac{1}{(\widetilde{E}_{\bm{k}A} + \widetilde{E}_{\bm{k}C})(\widetilde{E}_{\bm{k}B} + \widetilde{E}_{\bm{k}C})}\Bigg]\Bigg[ \frac{1}{(\widetilde{E}_{\bm{k}A} + \widetilde{E}_{\bm{k}C})^2} + \frac{1}{(\widetilde{E}_{\bm{k}B} + \widetilde{E}_{\bm{k}C})^2} \Bigg]\Bigg) \\
        & + \frac{\widetilde{U}^2_{\bm{k}AC}\widetilde{U}^2_{\bm{k}BC}}{(\widetilde{E}_{\bm{k}A} + \widetilde{E}_{\bm{k}B})^2}\Bigg(\Bigg[ \frac{1}{(\widetilde{E}_{\bm{k}A} + \widetilde{E}_{\bm{k}C})^3} + \frac{1}{(\widetilde{E}_{\bm{k}B} + \widetilde{E}_{\bm{k}C})^3} \Bigg]
        +
        \frac{1}{\widetilde{E}_{\bm{k}A} + \widetilde{E}_{\bm{k}B}}\Bigg[ \frac{1}{(\widetilde{E}_{\bm{k}A} + \widetilde{E}_{\bm{k}C})^2} + \frac{1}{(\widetilde{E}_{\bm{k}B} + \widetilde{E}_{\bm{k}C})^2} \Bigg]
        \Bigg)
        \Bigg\}.
    \end{split}
\label{eq:drag_fourth_order}
\end{equation}

\end{widetext}
The superscript indicates in which order in the perturbation expansion the expression originates, and the superscript $AB$ indicates that the drag is between the BEC components $A$ and $B$. The expressions for $\rho_{d,AC}$ and $\rho_{d,BC}$ are simply obtained by an interchange of indices. It is therefore sufficient to only consider $\rho_{d,AB}$.
\vspace{0.3cm}

\begin{figure}[!ht]
 	\centering
 	\includegraphics[scale=0.6]{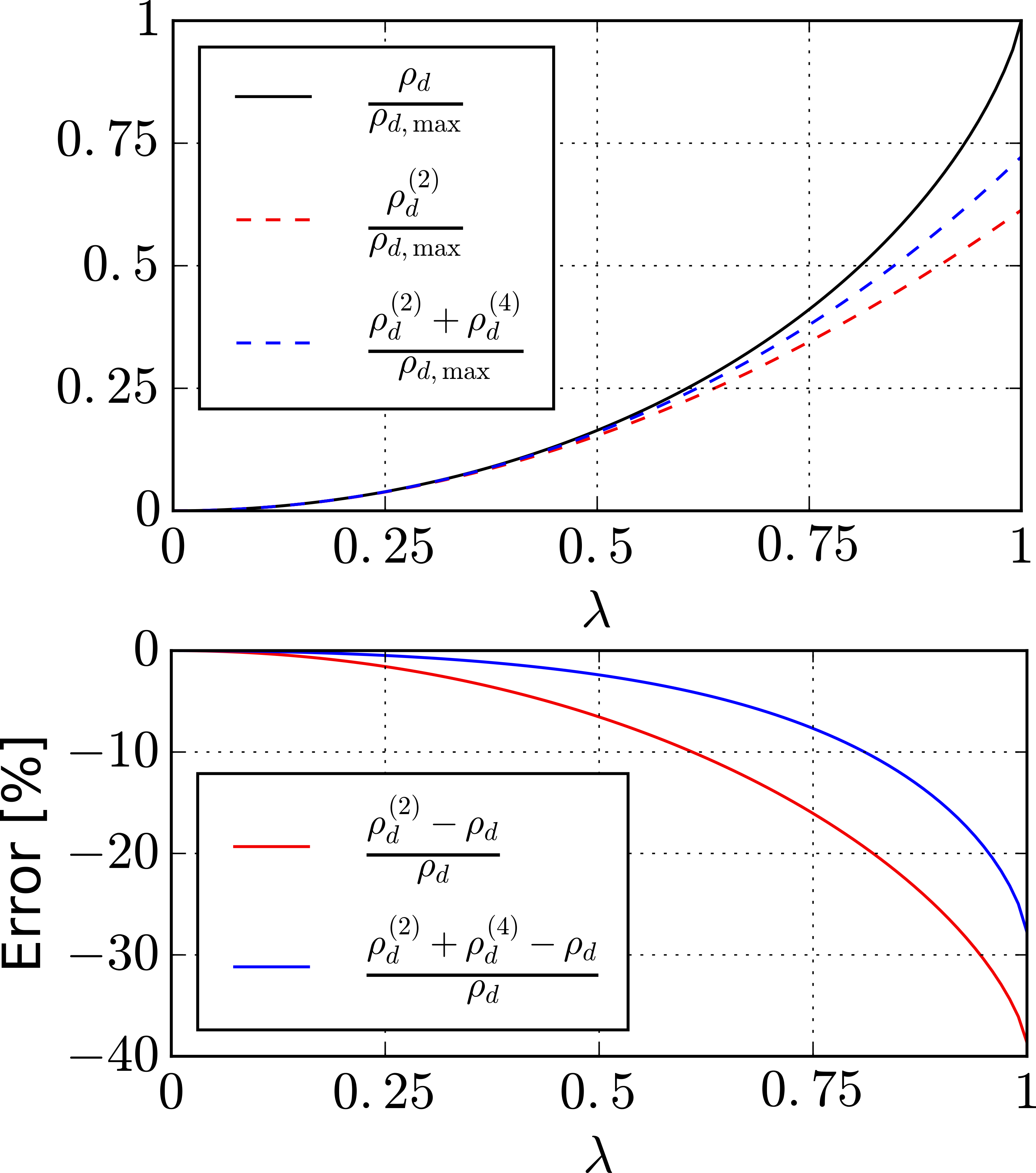}
 	\caption{The perturbation expansion \eqref{eq:drag_density_perturbation_full} for the superfluid drag density is compared to the exact expression \eqref{eq:superfluid_drag_density} in the two-component case, with $N_s=10^6$, $d=1$, $t_{\alpha}=1$, $m_{\alpha}=1$, $\gamma_{\alpha}=\gamma=1$, and $n_{\alpha} = 0.3$, while varying the inter-component interaction $\lambda_{AB}=\gamma_{AB}/\gamma$. Here, we have set $T=0$.}
 	\label{fig:Superfluid_drag_density_compare_2nd_and_4th_order_approx}
 \end{figure}

   \begin{figure}[!ht]
 	\centering
 	\includegraphics[scale=0.8]{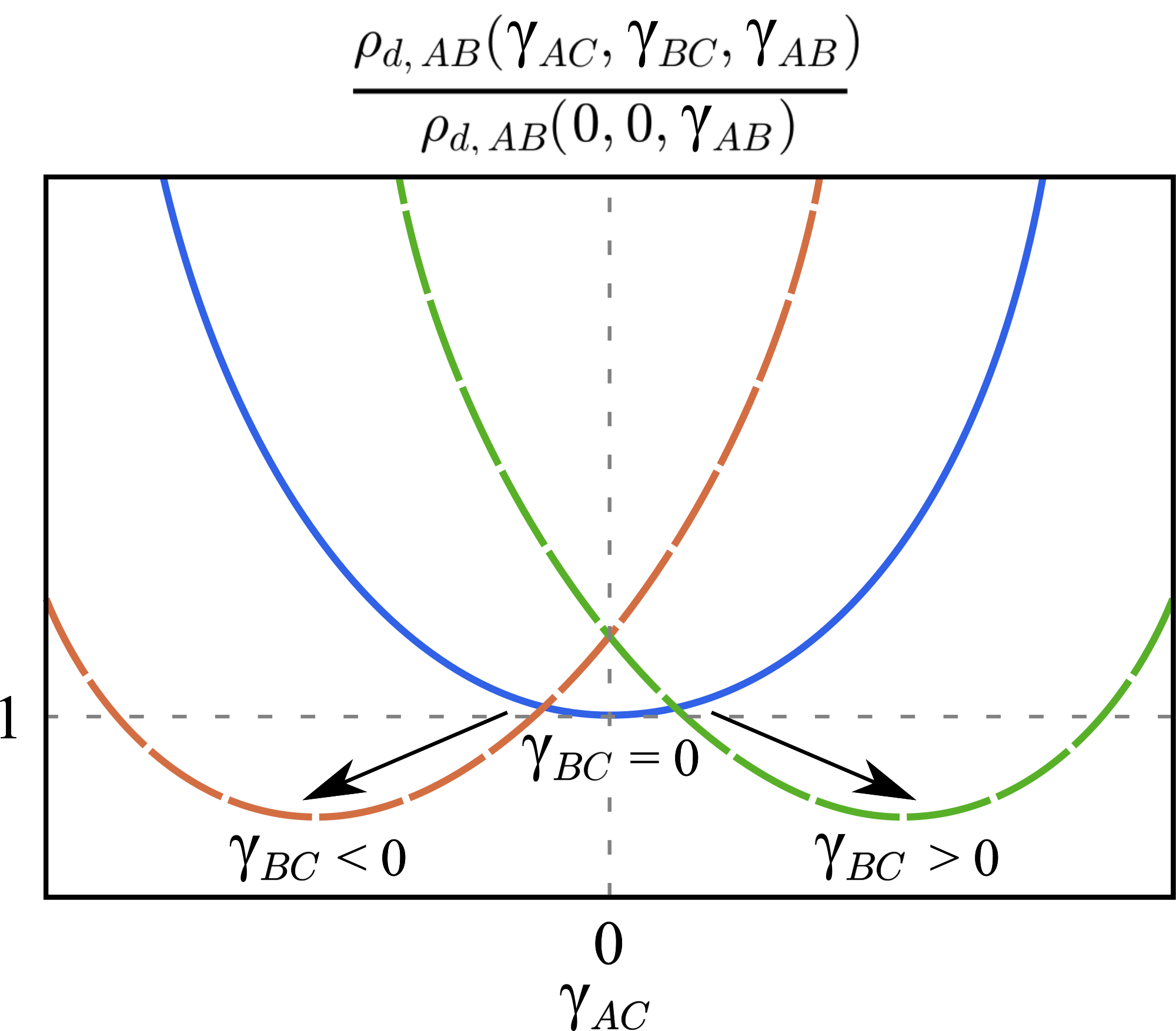}
 	\caption{A qualitative illustration of how the behaviour of the drag is shifted due to the introduction of a third component. When $\gamma_{AB}, \gamma_{AC}\neq 0$ and $\gamma_{BC}=0$ the drag originates from second- and fourth-order contributions and is parabolic with respect to $\gamma_{AC}$ with the minimum at the origin. As $\gamma_{BC}$ is turned on, the minimum is shifted down and to the side due to the third-order contribution, with direction depending on the sign of $\gamma_{BC}$. In the figure it is assumed that $\gamma_{AB}>0$, but with $\gamma_{AB}<0$ the direction of the sideways shift is reversed.
 	}
 	\label{fig:Third_Component_Effect_On_Drag}
 \end{figure}

A simple check on the usefulness of the perturbation expansion can be found by comparing \eqref{eq:drag_density_perturbation_full} to the exact expression \eqref{eq:superfluid_drag_density} for a two-component condensate.  This is done in \figref{fig:Superfluid_drag_density_compare_2nd_and_4th_order_approx}. In this case $\widetilde{U}_{\bm{k}\alpha\beta}=0$ and $\rho^{\text{(3)}}_{d,AB}=0$, and the subscript $AB$ is dropped to emphasize that there is only one drag-coefficient. (The exact two-component expression expanded to fourth order in the inter-component interaction between $A$ and $B$ yields the same result as above when the inter-component interactions between $A$ and $C$, and $B$ and $C$ is set to zero). It can be seen that the perturbative expression yields the same qualitative behaviour and is fairly accurate at small inter-component interactions.
 
 In the three-component case, the expansion \eqref{eq:drag_density_perturbation_full} provides qualitative insight into how the third boson component $C$ will affect the drag between $A$ and $B$.
 The third-order contribution implies that the drag can be enhanced or diminished, depending on the combination of strengths and signs of the inter-component interaction. From \eqref{eq:drag_third_order} the condition for enhancing the drag is $\gamma_{AB}\gamma_{AC}\gamma_{BC} < 0$, and  diminishing it is $\gamma_{AB}\gamma_{AC}\gamma_{BC} > 0$. However, this inequality is not precise due to fourth-order corrections. How the drag is influenced by introduction of the third-order contribution is displayed in figure \figref{fig:Third_Component_Effect_On_Drag}. The third component need, however, only couple to one of $A$ and $B$ to affect the drag between them, as seen from the fourth-order contribution \eqref{eq:drag_fourth_order}. In this case the drag increases and does not depend on the sign of the interactions. Even when the interaction between $A$ and $B$ vanishes, $\gamma_{AB}=0$, the third component can mediate the drag between them, resulting in a positive $\rho_{d,AB}$, as seen from the fourth-order contribution with $\widetilde{U}_{\bm{k}AB} = 0$. Second- and third-order contributions vanish in this case.
 
\begin{figure}[!h]
 	\centering
 	\includegraphics[scale=0.6]{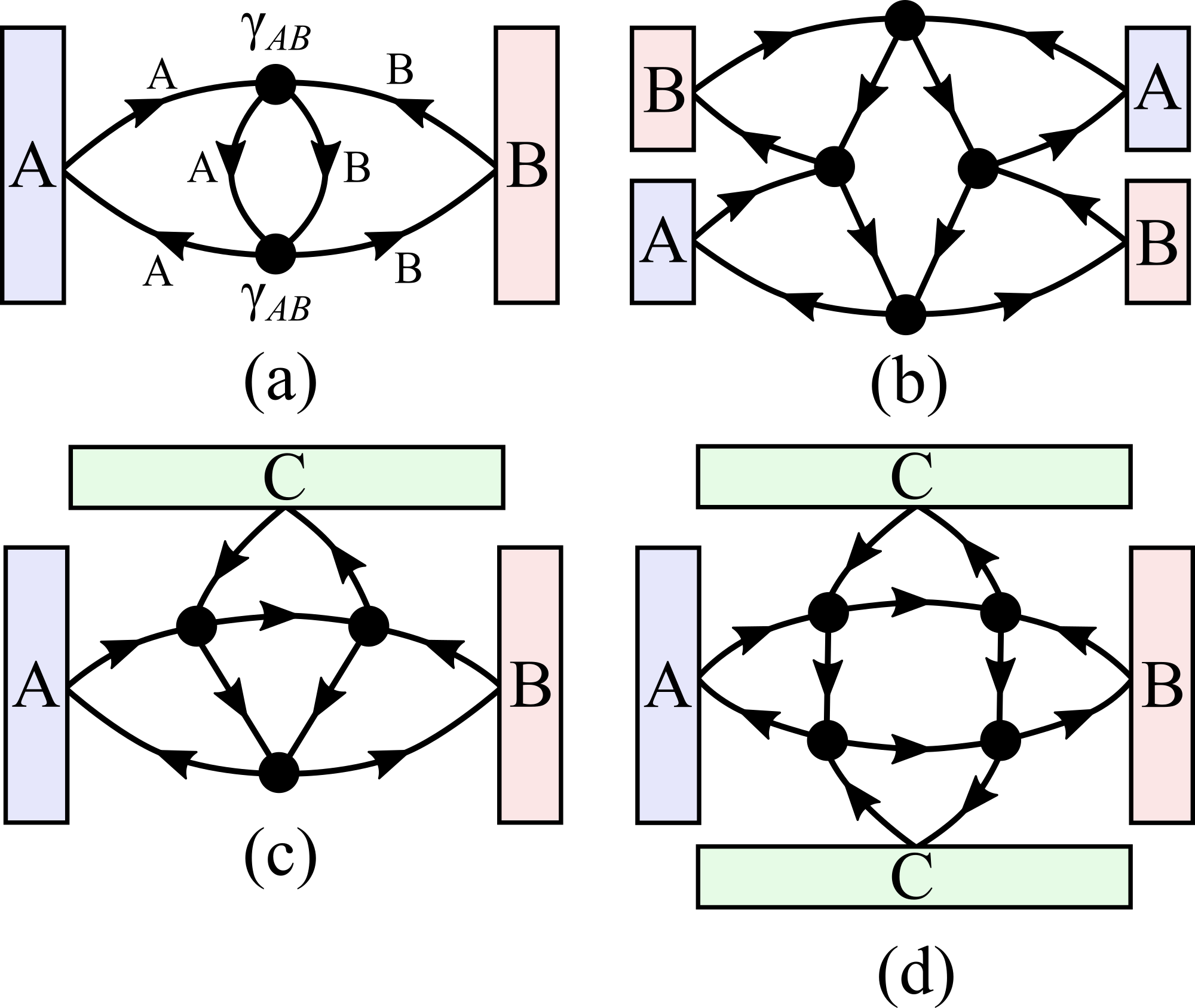}
 	\caption{The Feynman-diagrams for the perturbation expansion of the momentum-momentum correlation function yielding the superfluid drag coefficient, illustrating the Rayleigh-Schrödinger perturbation expansion as excitations moving in and out of the condensate and their interactions with one another. The large rectangles represent the components of the BEC which play the part of particle reservoirs, the arrowed lines the path of the excitations, and circled vertices the two-body interactions. The diagram "rules" are as follows: (1) Particle type and number is conserved at each vertex and must always include two distinct boson components moving in and two moving out. (2) Each vertex contributes a factor proportional to $\gamma_{\alpha\beta}$ where $\alpha$ and $\beta$ are the boson components connected to the vertex. (3) Each vertex has two lines connected to other vertices, and two lines connected to BEC reservoirs. (4) All the excitation paths are closed in the sense that each line exiting a BEC reservoir must return to the reservoir. Diagrams (a) and (b) illustrate the second- and fourth-order processes, respectively, contributing to the drag coefficient in a two-component condensate. Diagram (c) illustrates a third-order process in a three-component condensate, a process which has no counterpart in a two-component condensate. Diagram (d) illustrates one of several fourth order process possible in a three-component condensate.}
 	\label{fig:RS_Pert_Expansion_Diagrams}
 \end{figure}

 The perturbative expansion can be illustrated as excitations moving in and out of the condensate and their interactions, shown in \figref{fig:RS_Pert_Expansion_Diagrams}. Diagram (a) represents the second-order, (c) third-order, and (b) and (d) some of the fourth-order contributions. Specifically, (b) can be interpreted as a fourth-order term contributing to the usual two-component drag, while (d) is a fourth-order contribution to $\rho_{d,AB}$ where the drag is entirely mediated via the third component $C$, i.e.\ $A$ and $B$ do not interact directly. Diagram (c), a third-order term with no counterpart in the two-component case, and which may cause a reduction of the drag, illustrates a process where the condensates $A$ and $B$ partially interact directly and partially interact indirectly.
 
\begin{figure*}
 	\centering
 	\includegraphics[scale=0.65]{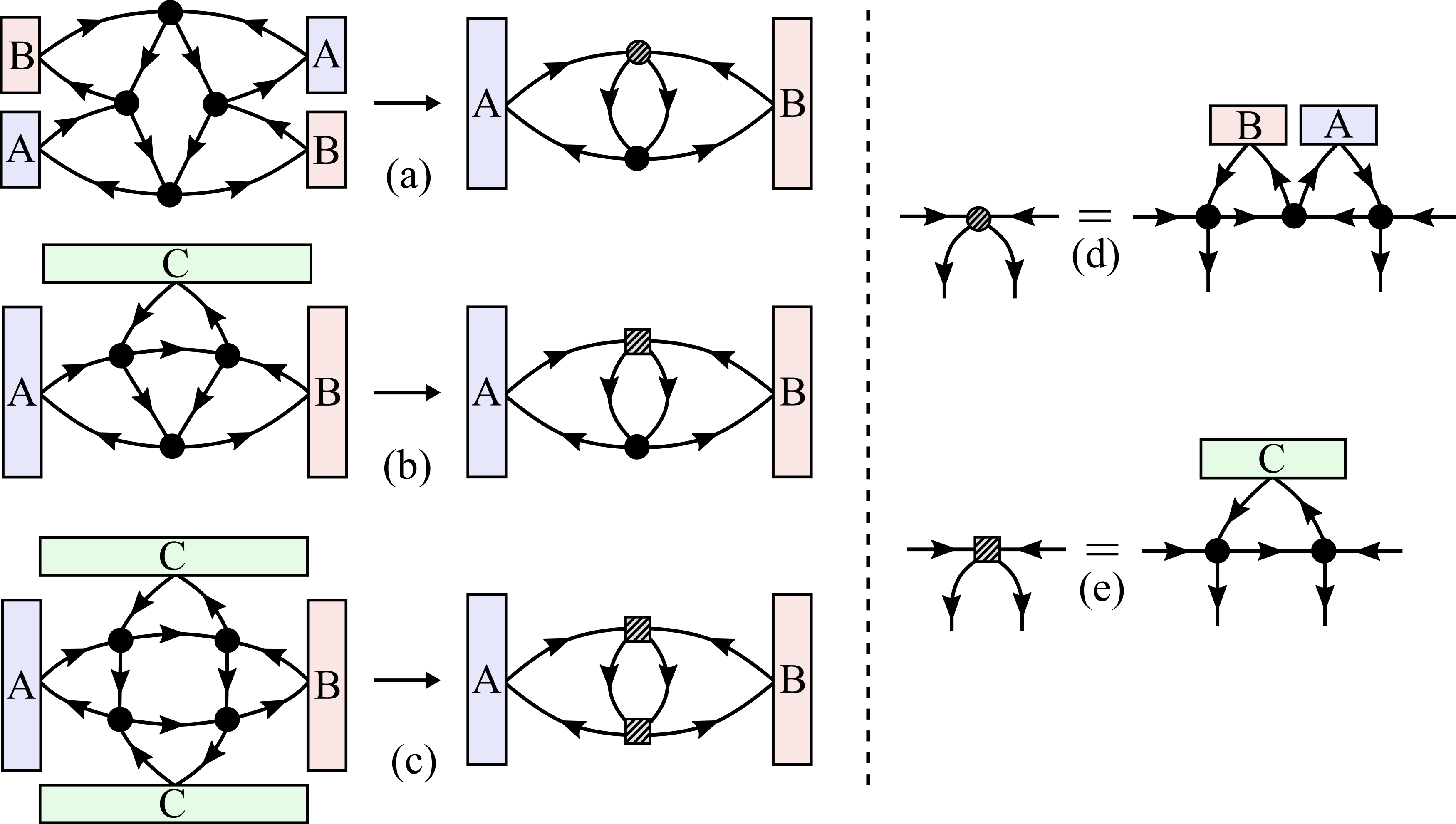}
 	\caption{The higher-order contributions in \figref{fig:RS_Pert_Expansion_Diagrams} are drawn as two-body "skeleton" diagrams with renormalized \textit{effective} interaction vertices. Diagrams (a), (b), and (c) are explained in Fig. \ref{fig:RS_Pert_Expansion_Diagrams}. Diagram (d) illustrates how a fourth-order process in a two-component system may be viewed as a second-order process with one renormalized interaction vertex involving three basic interactions. This renormalized vertex does not have a definite sign, but the remaining basic vertex (black dot) in  the diagram is of the same type as the three basic vertices of the renormalized vertex, and the resulting contribution to the superfluid drag coefficient is therefore always positive regardless of the sign of the inter-component interaction. Diagram (e) illustrates a renormalized vertex which does not have a counterpart in a two-component condensate.  The contribution to the superfluid drag coefficient $\rho_{d,AB}$ has a contribution from an interaction with the third-component, which also renormalizes a vertex in a "skeleton" diagram for $\rho_{d,AB}$. This vertex has a magnitude  given by $\gamma_{AC} \gamma_{BC}$, which is independent of the basic vertex $\gamma_{AB}$ entering the contribution. This contribution to the superfluid drag $\rho_{d,AB}$ (a contribution that does not exist in a two-component case) can therefore be either positive or negative. }
 	\label{fig:RS_Pert_Expansion_Effective_Diagrams}
 \end{figure*}

  In \figref{fig:RS_Pert_Expansion_Effective_Diagrams} the higher-order diagrams of \figref{fig:RS_Pert_Expansion_Diagrams}, which correspond to three- and four-body collisions, are drawn as effective second-order "skeleton" diagrams displaying interaction between components $A$ and $B$. In this way, the drag is understood as arising from collision events where increasingly complicated interaction mechanisms, in which bosons of all components move in and out of the BEC in the intermediate states, are renormalized into \textit{effective} two-component interactions.
 
 Note that, in the weak-coupling limit, the drag $\rho_{d,AB}$ is mainly determined by second-order collisions between particles of type A and B. As we have seen, there are important third- and fourth-order 
 {\it corrections} to this involving collisions between three {\it different types of particles}. This phenomenon has no analog in a two-component Bose-Einstein condensate. 
 For small $\gamma_{AC}/\gamma_{AB}$, the dominant correction to the second-order term  (the latter being also present in the two-component case) is the cubic term involving all interactions $\gamma_{AB}, \gamma_{AC}, \gamma_{BC}$, which does not have a definite sign. The third-order collision term explains the initial linear in $\gamma_{AC}/\gamma_{AB}$ correction to 
 $\rho_{d AB}$, provided $\gamma_{BC} \neq 0$.  
 Eventually, as $\gamma_{AC}/\gamma_{AB}$ increases, fourth-order terms with a definite sign in the interactions (products of squares) will dominate. 
 
 Returning to the numerical mean-field results of the previous subsection \ref{SubsectionA}, we see that 
 the numerical results shown in \figref{fig:Three_Component_Drag_Numerical} of that subsection are entirely in accord with the above considerations extracted from Eqs. \eqref{eq:drag_second_order} - 
 \eqref{eq:drag_fourth_order}. In particular, the sign of the {\it corrections} to the direct term $\sim \lambda_{AB}^2$ as a function of $\lambda_{AC}$ 
 and $\lambda_{BC}$ as well as the evenness of these corrections under a simultaneous change of the signs of $\lambda_{AC}$  and $\lambda_{BC}$, now become clear.
 
 This picture generalizes to systems with more than three distinct components of the condensate. For instance, in a four-component system there are $6$ inter-component interactions 
 $\gamma_{AB}, \gamma_{AC}, \gamma_{AD}, \gamma_{BC}, \gamma_{BD}, \gamma_{CD}$. 
 The expression for $\rho_{d,AB}$ therefore may be organized in a power series involving a quadratic term $\gamma_{AB}^2$, and leading order correction terms to this terms ($60$ in all) involving one factor of $\gamma_{AB}$ and two factors of the five other remaining interactions. 
 
\subsection{Beyond Weak-Coupling Mean-Field Theory: Quantum Monte Carlo Simulations}\label{SubsectionC}
In this subsection, we present Quantum Monte Carlo results for the drag density $\rho_{d,AB}$. This approach goes beyond mean-field theory, taking all fluctuations of the quantum fields into account, and also beyond all orders in perturbation theory. As such, it should yield useful insight into the validity of the results presented in the previous two subsections. 

For the path integral Quantum Monte Carlo results for $\rho_{d,AB}$ that we present in this paper, we start
directly from the Bose-Hubbard Hamiltonian 
\eqref{B-H}-\eqref{B-H-2}. The partition function can then be expressed in the imaginary time path integral formalism, projecting the quantum system onto a $(D+1)$-dimensional system described by classical configurations. The configurations consist of sets of worldlines that represent periodic particle trajectories in the extra dimension, namely imaginary time \cite{SFSOM}. Average values can then be obtained through Monte Carlo sampling. The statistics of the worldlines are related to the superfluid density through the Pollock-Ceperley formula \cite{PC1987}, which in the single-component case takes the form

\begin{equation}
    \rho_s = \frac{m^2 L^{2-D}}{D\beta}\langle W^2 \rangle,
\end{equation}

where $\bm{W}$ is the $D$-dimensional winding number vector, $\langle \,\,\,\, \rangle$ represents the average and $L = (N_s)^{1/D}$ is the number of lattice sites per dimension. The winding number $W_i$ is a a topological quantity, for a system with periodic spatial boundary conditions, counting the net number of times the particle trajectories wind around the system in direction $i$. In the multicomponent case, following the derivation of Ref \cite{Babaev2018}, the drag is given by

\begin{equation}
    \rho_{d, AB} = \frac{m_A m_B L^{2-D}}{D\beta}\langle \bm{W}_A \cdot \bm{W}_B \rangle,
\end{equation}
where $\bm{W}_A$ is the winding number vector for particles of component $A$.
\newpage

    \begin{figure}[!htb]
 	\centering
 	\includegraphics[scale=0.8]{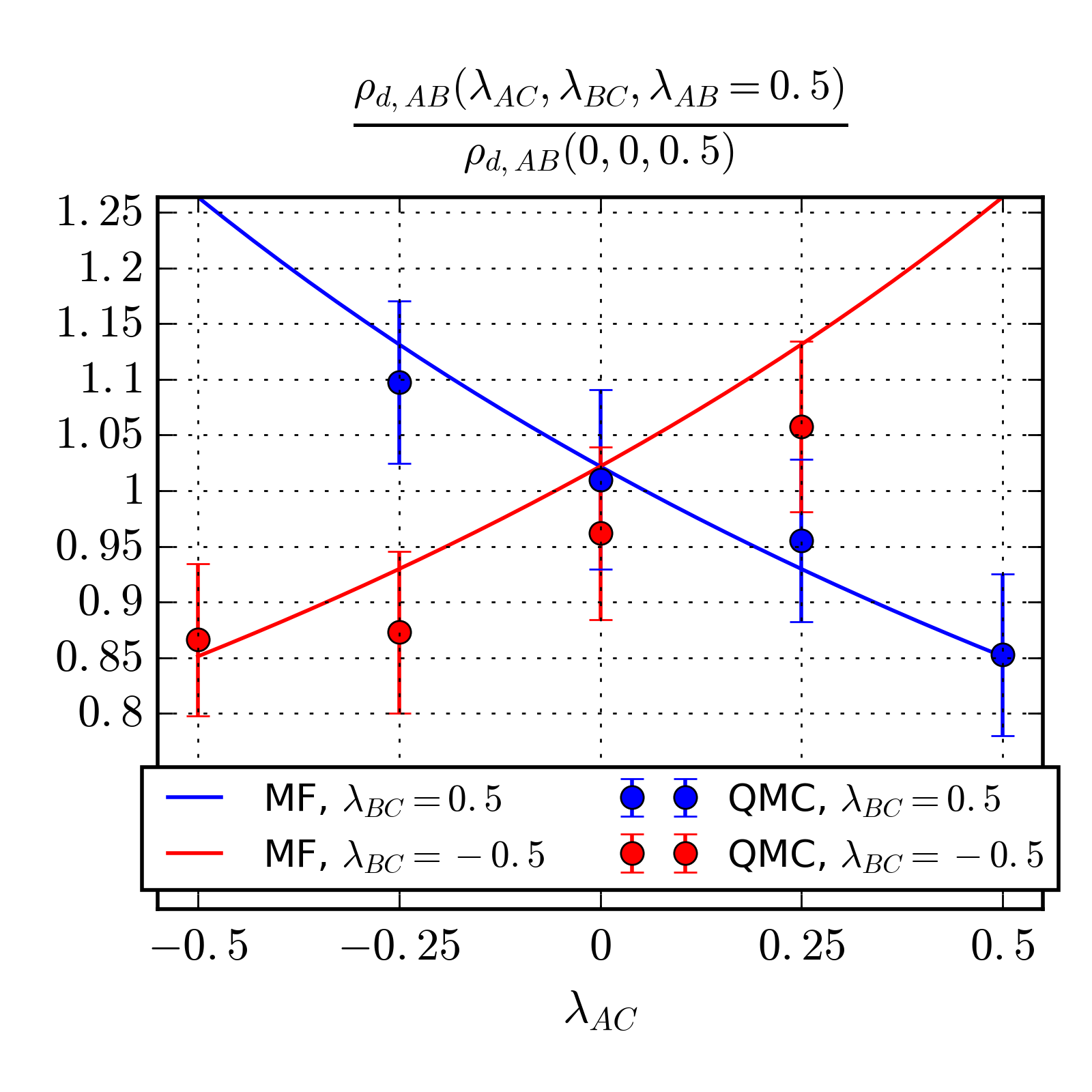}
 	\caption{The effect of a third component $C$ on $\rho_{d,AB}$ is computed from mean-field theory (MF) and Quantum Monte Carlo (QMC) simulations with $T = 0.1$,  $N_s = 100$, $d=1$, $t_{\alpha}=1$, $\gamma_{\alpha} = \gamma = 1$, $m_{\alpha} = 1$, and $n_{\alpha}=0.3$ with $\lambda_{\alpha\beta}\equiv\gamma_{\alpha\beta}/\gamma$. The mean-field results correspond to two of the curves from figure \figref{fig:Three_Component_Drag_Numerical} (d), apart from the difference in temperature and lattice size.}
 	\label{fig:Drag_MF_VS_QMC}
 \end{figure}

We present results obtained using the worm algorithm, which is well-suited for sampling configurations with different winding numbers \cite{Prokofev1998, Prokofev1998v2}. Values for the drag for fixed particle numbers were achieved by tuning the chemical potentials to produce the desired average particle numbers. Comparison between results from Monte Carlo simulations and mean-field theory are displayed in \figref{fig:Drag_MF_VS_QMC}. The error bars have been obtained through bootstrapping \cite{Efron1986} and include both the error in the nominator and the denominator. The Quantum Monte Carlo simulations produced relative results in accordance with mean-field theory, but predicts the normalization value $\rho_{d,AB}(0, 0, 0.5)$ to be about $40\%$ higher. The qualitative behavior obtained from mean-field theory is clearly supported by the Monte Carlo results.

This lends further support to the picture we have presented in \ref{SubsectionB}, in which the decrease of the drag as a function of $|\lambda_{AC}|$, provided that $\lambda_{AC} \lambda_{BC} > 0$, is explained.

\subsection{Finite-Temperature Mean-field Results}\label{SubsectionD}

The Monte Carlo results of Section \ref{SubsectionC} are obtained at a temperature $T=0.1$ in units where $t_{\alpha} = \gamma = 1$, while the numerical and analytical mean-field results of subsections \ref{SubsectionA} and \ref{SubsectionB} are obtained at $T=0$. However, since we expect the transition to the normal state to take place at a temperature $T \sim {\cal{O}}(1) \gg 0.1$, we expect a comparison between the $T=0$-results of   subsections \ref{SubsectionA} and \ref{SubsectionB} on the one hand, and subsection \ref{SubsectionC} on the other hand, to be meaningful. 

We next substantiate this claim by presenting numerical mean-field finite-T results for the drag-coefficients. 
 At finite temperatures the free energy is
 
 \begin{equation}
     \begin{split}
         \mathcal{F} = &\frac{\widetilde{H}_0}{N_s} - \frac{1}{2N_s}\sum_{\bm{k}\neq 0}[E_{\bm{k}A} + E_{\bm{k}B} + E_{\bm{k}C}] + \frac{1}{4N_s}\sum_{\bm{k}\neq 0}\sum_{i=1}^{6}\mathcal{E}_{\bm{k}i}\\
         &\quad + \frac{1}{2\beta N_s}\sum_{\bm{k}\neq 0}\sum_{i=1}^{6}\ln\big(1-e^{-\beta\mathcal{E}_{\bm{k}i}}\big).
     \end{split}
 \end{equation}
 
As mentioned in appendix \ref{app:Diagonalization}, $\mathcal{E}_{\bm{k}i}$ yields three distinct bands, not six. However, for convenience during computations the spectrum is treated as six bands, which is the reason for including the $1/2$ factor in front of the temperature dependent part of the free energy.
 
\begin{widetext}
 
 \begin{figure}[H]
 	\centering
 	\includegraphics[scale=0.63]{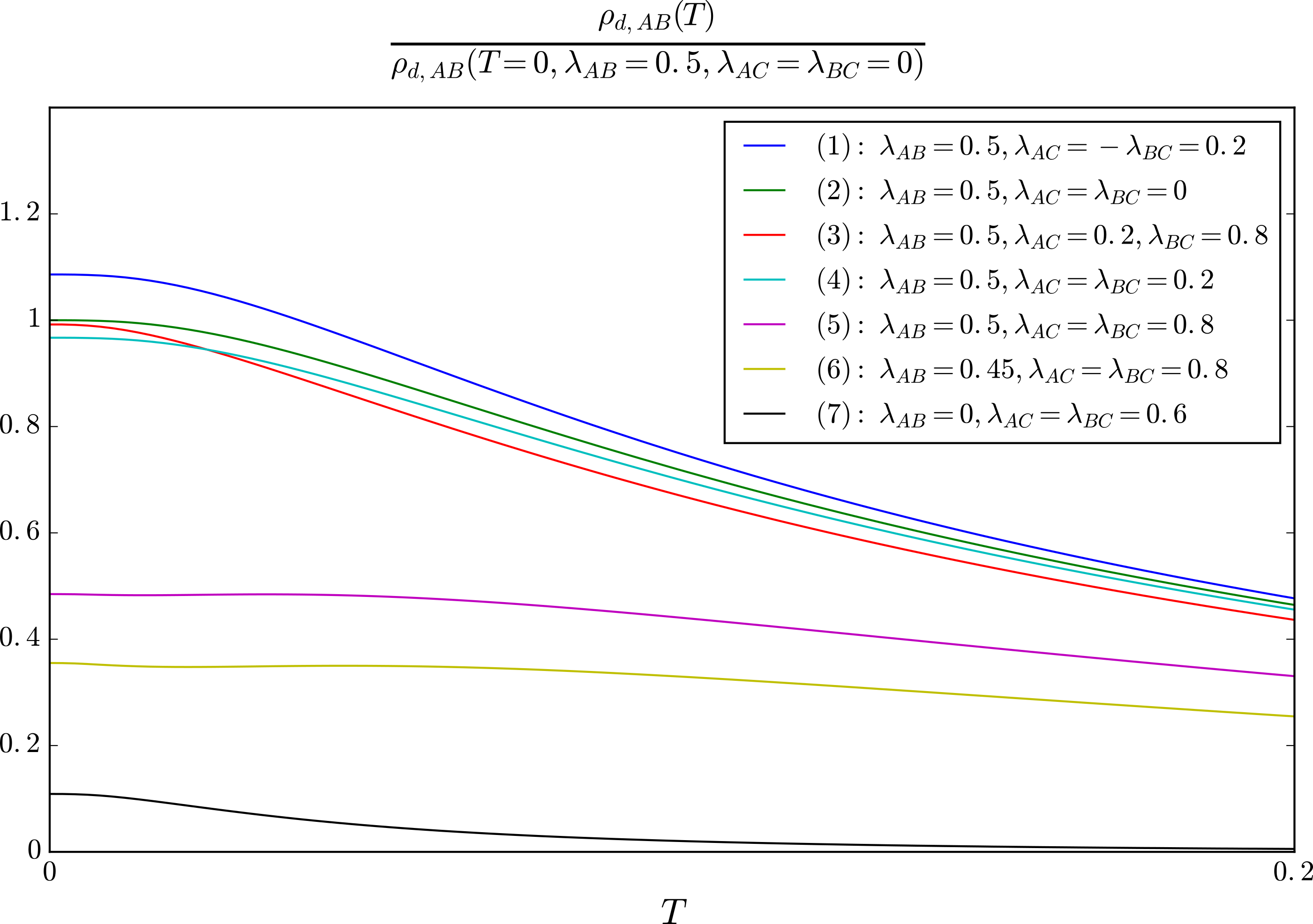}
 	\caption{The temperature dependence of $\rho_{d,AB}$ is shown at various combinations of the inter-component interaction strengths, normalized by the value at $T=0$, $\lambda_{AB}=0.5$, and $\lambda_{AC}=\lambda_{BC}=0$. The plots are numbered by decreasing value at $T=0$. The parameters used are  $N_s = 10^4$, $d=1$, $\Delta = 2\pi 10^{-4}$, $t_{\alpha}=1$, $m_{\alpha}=1$, $n_\alpha=0.3$, $\gamma_{\alpha} = \gamma = 1$ for $\alpha = $A, B and C. All the plots go towards zero in the same manner at high $T$, not shown here. }
 	\label{fig:Three_Component_Drag_Finite_T}
\end{figure}
\end{widetext}

 The temperature dependence of the superfluid drag density is shown in \figref{fig:Three_Component_Drag_Finite_T} at various combinations of the inter-component interaction strengths up to a temperature of $T=0.2$. This is twice the temperature that was used in the Quantum Monte Carlo simulations in the previous section. We see that in all cases apart from the case (7), where the drag-coefficient is the smallest, the relative reduction of $\rho_{d,AB}$ is modest over the entire span of temperatures $T \in[0.0,0.2]$ and certainly in the range $T \in[0.0,0.1]$.    
 One point to note is that three of the plots show behavior that differ from the rest. In the curve labelled by (3), the drag initially decreases faster than the others, even falling below the curve labelled by (4) which has a smaller value at $T=0$. The drag in (5) and (6) remains almost constant as a function of $T$ for small temperatures $T \leq 0.1$, although the curve labelled by (6) has a slight initial decrease at very small $T$. The main conclusion to be drawn from  this is that the thermal suppression of $\rho_{d,AB}$ in this temperature-range is very modest, which means that computations carried out at $T=0$ give reasonably accurate results for $\rho_{d,AB}$ at $T=0.1$.
 
In the previous sections we have seen how we, at the mean-field level, can express $\rho_{d,AB}$ as sums over momentum space. In this way, we may view the total drag-coefficient (and indeed any other component of the superfluid tensor), as a sum over Fourier-modes of the superfluid tensor, on the following form
\begin{eqnarray}
\rho_{dAB} = \frac{1}{N_s} 
\sum_{\bm{k} \neq 0} ~ \rho_{dAB} (\bm{k} ), 
\end{eqnarray}
as in Eqs. \eqref{eq:superfluid_drag_density} and
\eqref{eq:drag_second_order}-\eqref{eq:drag_fourth_order}. \footnote{In the $\bm{k}$-summation, the
$\bm{k}=0$-term has been, and should be, omitted. Due to the factors $\sin^2 (k_x)$ entering in all cases, this omission is not necessary.}

It is therefore instructive and of some interest to also consider the $T$-dependence of th distribution of $\rho_{d,AB}$ in $\bm{k}$-space, since it yields insights into which quasiparticle excitations  are mainly responsible for depleting  $\rho_{d,AB}$. 

In \figref{fig:Three_Component_Drag_Finite_T_momentum_distribution} and \figref{fig:Three_Component_Drag_Finite_T_momentum_distribution_extra} we show the distribution of $\rho_{d,AB}$ in  $\bm{k}$-space  for various coupling constants and temperatures. The parameters are the same as in \figref{fig:Three_Component_Drag_Finite_T}, but with $N_s = 200^2$ for increased resolution in $\bm{k}$-space. From the figure, it is seen that the temperature dependence can be understood as the suppression of contributions to $\rho_{d,AB}$ in an increasing region around the origin in $\bm{k}$-space. This explains why the curve labelled (3) in \figref{fig:Three_Component_Drag_Finite_T} initially decreases faster with $T$ than the curve labelled by (5). Namely, the majority of the drag in (3) comes from momenta very close to the origin, while in (5) the main contribution is at larger momenta and survives longer as the region of suppression is increased with $T$. Thermally exciting long-wavelength quasiparticles, thereby depleting all components of the superfluid density including the superfluid drag, costs less energy at low $\bm{k}$ than larger $\bm{k}$ and thus the Fourier-modes of the superfluid densities at small momenta are more susceptible to thermal depletion than at larger momenta.  

The decrease in drag as the temperature increases is seen to depend on the inter-component interactions, and can be understood as the suppression of contributions in an increasing region around the origin in $\bm{k}$-space. The majority of the drag in $\bm{k}$-space is distributed as a symmetric pair of lobes around the origin. This is also the case for the two-component
case, and the lobe-structure of the momentum-distribution of $\rho_{d,AB}$ can therefore be understood from Eq. \eqref{eq:superfluid_drag_density}. The bare dispersion relations $\epsilon_{\bm{k} \alpha}$ are quadratic in $k$ for small $\bm{k}$, while 
$\mathcal{E}_{\bm{k}\pm}$ in Eq. \eqref{eq:two_component_eigenenergy} are essentially the multi-component versions of the
Bogoliubov-spectrum which is linear in $k$. The momentum-distribution of $\rho_{d,AB}$ therefore increases linearly with due to the factor $\sin^2(k_x)$. At larger $k$, the momentum distribution essentially varies as 
$1/(\mathcal{E}_{\bm{k} +} + \mathcal{E}_{\bm{k} -})^3$ and therefore drops off. In addition, at the zone-edge
$k_x = \pm \pi$, the factor $\sin^2(k_x)$ vanishes. \footnote{The factor $\sin^2(k_x)$ is due to our choice of twist-vector along the $\hat x$-axis. Had we chosen the twist-vector along the $\hat y$-axis, the factor would have been $\sin^2(k_y)$ and the lobes would have been located along the $\hat y$-axis.}

\begin{widetext}

\begin{figure}[H]
 	\centering
 	\includegraphics[scale=1.2]{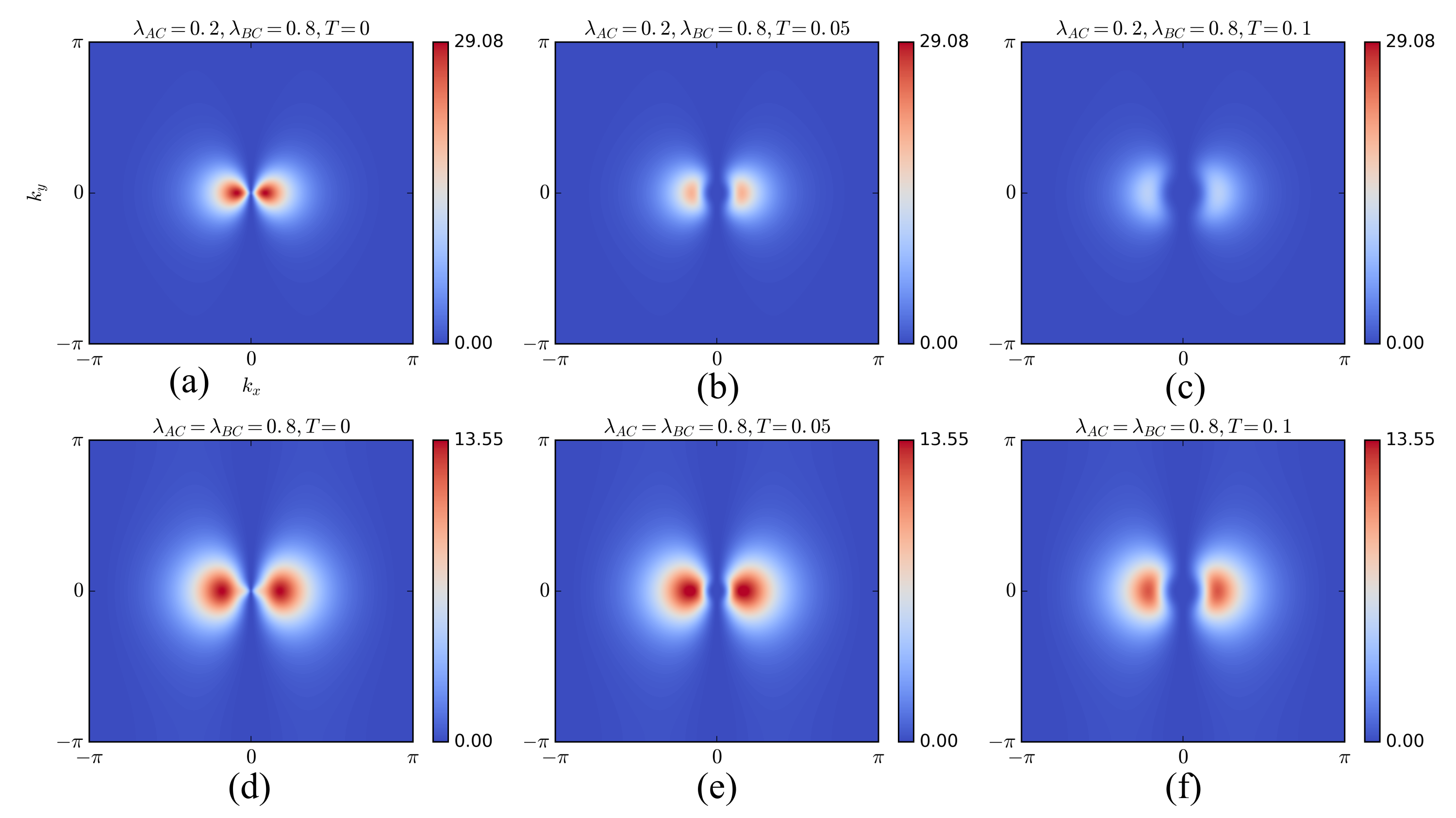}
 	\caption{The contribution to the drag density as a function of momentum is shown for parameters corresponding to the curve labelled by (3) in (a)-(c), and parameters corresponding to the curve labelled by (5) in (d)-(f), of \figref{fig:Three_Component_Drag_Finite_T} at various temperatures in units of the mean contribution per $\bm{k}$-vector at $T=0$; $\rho_{d,AB}(T=0)/N_s$. The parameters are the same as in \figref{fig:Three_Component_Drag_Finite_T}, but with $N_s = 200^2$ for increased resolution in the $\bm{k}$-plane. Note how the region of $\bm{k}$-space where the momentum-distribution of the drag is suppressed around $\bm{k}=0$ increases with $T$. The dominant contribution to the suppression of the superfluid drag comes from long-wavelength quasiparticles, which are energetically least costly to thermally excite out of the condensate.}
 	\label{fig:Three_Component_Drag_Finite_T_momentum_distribution}
\end{figure}

\begin{figure}[H]
 	\centering
 	\includegraphics[scale=1.2]{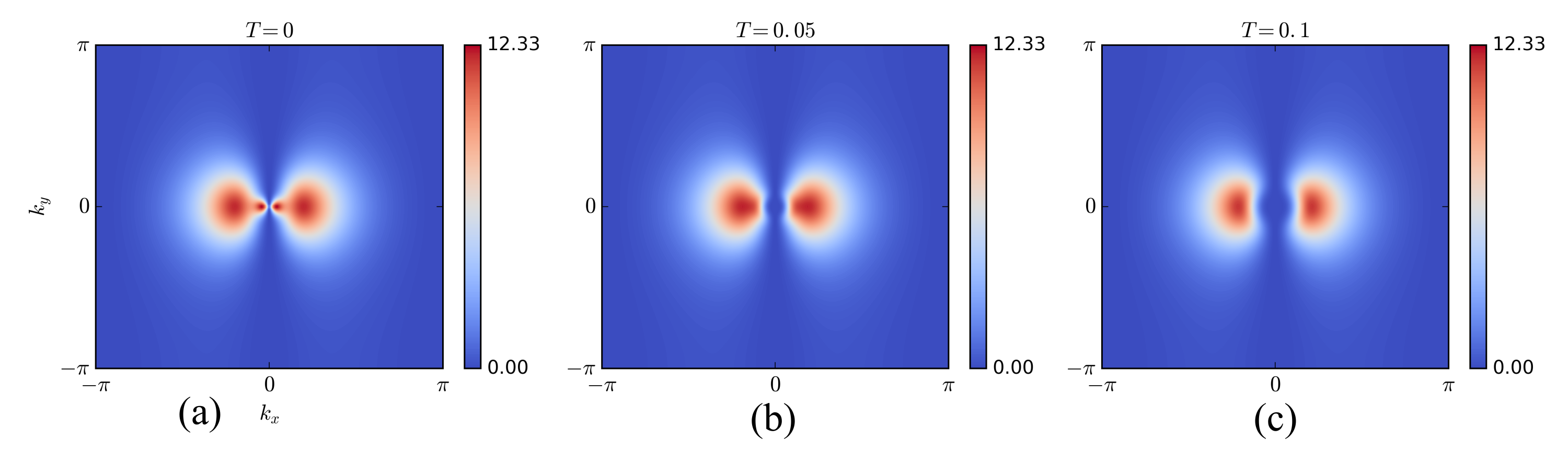}
 	\caption{The contribution to the drag density as a function of momentum is shown for parameters corresponding to the curve labelled by (6) of \figref{fig:Three_Component_Drag_Finite_T} at various temperatures, in units of the mean contribution per $\bm{k}$-vector at $T=0$; $\rho_{d,AB}(T=0)/N_s$. The parameters are the same as in \figref{fig:Three_Component_Drag_Finite_T}, but with $N_s = 200^2$ for increased resolution in the $\bm{k}$-plane. The distribution of drag in the $\bm{k}$-plane is seen to be a combination of the situation corresponding to the cases (3) and (5) in \figref{fig:Three_Component_Drag_Finite_T_momentum_distribution}: There are two inner and two outer lobes where the contribution is centered. Note how the region of $\bm{k}$-space where the momentum-distribution of the drag is suppressed around $\bm{k}=0$ increases with $T$. The dominant contribution to the suppression of the superfluid drag comes from long-wavelength quasiparticles, which are energetically least costly to thermally excite out of the condensate. }
 	\label{fig:Three_Component_Drag_Finite_T_momentum_distribution_extra}
\end{figure}
\end{widetext}

Varying interaction strengths changes the proximity of the lobes to the origin and explains the various temperature dependencies. The drag decreases faster with temperature when the lobes are close the origin and slower when they are further away. The same is true for the two-component case. However, in the three-component case an additional pair of lobes can emerge, as in \figref{fig:Three_Component_Drag_Finite_T_momentum_distribution_extra}. This causes the drag to experience an initial decrease followed by a plateau-like region. This feature is absent for two-component systems. While we have no closed expression for $\rho_{d,AB}$ analogous to Eq. \eqref{eq:superfluid_drag_density} for the three-component case, the perturbation expansion for $\rho_{d,AB}$ shows that its momentum-distribution in the three-component case is endowed with considerably more $\bm{k}$-space structure than the perturbation series in two-component case. The latter is obtained from the perturbation series Eqs. \eqref{eq:drag_second_order} - \eqref{eq:drag_fourth_order} for the three-component case by setting $U_{AC}=U_{BC}=0$. This accounts for the extra tiny lobes of significant values of the $\bm{k}$-space resolved $\rho_{d,AB}$ very close to $k=0$ in the three-component case.

\section{Summary}\label{Conclusion}

In conclusion, we have computed the inter-component drag-coefficients of a three-component Bose-Einstein condensate in the weak-coupling limit, well into the superfluid phase. We find a non-monotonic inter-component drag as a function of the inter-component interaction coefficients. The results are obtained analytically as well as numerically from the mean-field free energy of the system, and the results are confirmed by large-scale path integral Monte Carlo simulations using the worm-algorithm. The non-monotonicity we find has no counterpart in the two-component case, where drag increases monotonically with increasing inter-component interaction, regardless of the sign of this inter-component interaction. 

In the present case, the situation is considerably more complicated for three components, but reduces to what we expect based on the two-component case if one of the inter-component interactions is absent. 
For large inter-component interactions, the drag-coefficients are monotonically increasing functions of inter-component interactions, regardless of the sign of the interaction. However, for relatively small to intermediate inter-component interactions,  the drag-coefficient between component A and B decreases as a function of the inter-component interaction between A and a third component C for a repulsive fixed inter-component interaction between B and C. The situation is reversed for an attractive inter-component interaction between B and C. The drag-coefficients are all positive throughout in this superfluid regime. Therefore,  these results are quite different from the non-monotonic behavior one observes close to the Mott insulator transition, where drag-coefficients become negative due to strong correlations and backflow. We have identified the renormalization of the inter-component scattering vertex 
via a third-order term involving {\it three different particles, one particle being a virtual excitation in the third condensate component} as the origin of the non-monotonicity of the inter-component drag. This third-order renormalization may endow the bare and renormalized inter-particle scattering vertex with different signs. Fourth-order terms, which always have a positive sign, will eventually dominate the third-order term, leading to the naively expected increase in drag with increasing inter-component interaction. 

We attribute these {\it corrections} to the two-component results in the superfluid drag between two components to subtle three-body collisions which dominate four-body collisions at low values of $\gamma_{AC}/\gamma_{AB}$. These three-body corrections have no counterpart in two-component BECs. They are determined by the factor $\gamma_{AB} \gamma_{BC} \gamma_{AC}$ which may be positive or negative, depending on the sign of the interactions. In contrast, the second-order collision term determining $\rho_{d,AB}$ is determined by $\gamma_{AB}^2$ and always gives an increasing drag as a function of inter-component interaction. In principle, three-component BECs well into the superfluid phase therefore form a laboratory for investigating particle-interactions beyond standard pairwise interactions. 

We finally remark that we in this paper have focused on a model Hamiltonian without component-mixing interactions, \eqref{B-H}-\eqref{B-H-2}, describing a system of $N$ conserved components, such as a heterogeneous mixture of different atoms. For homo-nuclear mixtures where the different components consist of atoms of the same type, but in different hyperfine spin states, it could be necessary to also take into account scattering processes where the internal states of the atoms are altered during the processes. Of special interest for the three-component case are spinor-one systems where the atoms occupy three hyperfine spin states that constitute a complete hyperfine multiplet. An example of such a system is a $\rm{Rb}^{87}$-mixture with atoms prepared in three different hyperfine spin states with hyperfine spin quantum numbers $F=1$, $m_F=-1,0,1$. A model describing such a system would, in addition to the pure density-density interaction terms, also include interaction terms where e.g.\ two atoms with $m_F = 1$ and $m_F = -1$, respectively, produce two atoms with $m_F=0$. Such non-conservation of components vastly complicates the description. At the mean-field level, the theoretical complication is similar in character to the severe complication one faces when introducing synthetic spin-orbit coupling \cite{Spielman2011} into the description of a multi-component Bose-Einstein condensate. 


\begin{acknowledgments}
The implementation of the algorithm used in the Monte Carlo simulations was based on code provided by the ALPS project (\url{http://alps.comp-phys.org/}) \cite{Bauer2011, Albuquerque2007} and E. Thingstad. This research was supported in part with computational resources at NTNU provided by NOTUR. A.S. was supported by the Research Council of Norway through Grant Number 250985, "Fundamentals of Low-dissipative Topological Matter" as well as through a Center of Excellence Research Grant from the Research Council of Norway, Grant Number 262633 "Center for Quantum Spintronics".
\end{acknowledgments}

\appendix

\section{Computation of Path Integral}
\label{app:Computation_of_Path_Integral}
To compute the path integral \eqref{eq:path_integral_formulation_partition_function} some manipulations are needed. First the $k$-sum is taken to run over half of $\bm{k}$-space, so that the $-\bm{k}$ terms become explicit. A Fourier transformation of the fields into frequency space is then performed, $\phi_{\bm{k}\alpha}(\tau) = \beta^{-\frac{1}{2}}\sum_{n}\phi_{\bm{k}\alpha n} e^{-i\omega_{n}\tau}$, where $\omega_{n}=2\pi n/\beta$ with $n \in \{0,\pm 1, \pm 2,\ldots\}$ are Matsubara frequencies. This results in the replacement of the integral $\int^{\beta}_{0}\diff\tau \rightarrow \sum_{n}$, the derivatives $\partial_{\tau} \rightarrow \pm i\omega_{n}$, and the fields $\phi^{*}_{\bm{k}\alpha}\phi_{\bm{k'}\beta} \rightarrow \phi^{*}_{\bm{k}\alpha n}\phi_{\bm{k'}\beta n}$, $\phi_{\bm{k}\alpha}\phi_{\bm{k'}\beta} \rightarrow \phi_{\bm{k}\alpha n}\phi_{\bm{k'}\beta -n}$, etc. The $n$-sum is recast to run over $n \geq 0$ so that the $-n$ terms also become explicit. A factor $(1-\frac{1}{2}\delta_{n,0})$ is included to avoid counting $n=0$ twice. Finally, the column vector
 \begin{equation}
    \begin{split}
        \Phi_{\bm{k}n} = \big(&\phi_{\bm{k}A n},\, \phi_{\bm{k}B n},\, \phi^*_{\bm{-k}A n},\, \phi^*_{\bm{-k}B n} ,\\ \,
        &\phi_{\bm{k}A -n},\, \phi_{\bm{k}B -n},\, \phi^*_{\bm{-k}A -n},\, \phi^*_{\bm{-k}B -n}\big)^{\textrm{T}},
    \end{split}
 \end{equation}
is defined, so that
\begin{equation}
    S = S_0 + \sideset{}{'}\sum_{\bm{k}\neq 0,n}\Phi^{\dagger}_{\bm{k}n} \mathcal{M}_{\bm{k}n} \Phi_{\bm{k}n},
\end{equation}
where $S_0$ are the terms that do not depend on the fields, and
\begin{equation}
\mathcal{M}_{\bm{k}n} = (1-\frac{1}{2}\delta_{n,0})
	\begin{bmatrix}
 		\mathcal{N}^{+}_{\bm{k}n} & \mathcal{N}_{\bm{k} n} \\
 		\mathcal{N}_{\bm{k}n} & \mathcal{N}^{-}_{\bm{k} n} \\
	\end{bmatrix},
\end{equation}

\begin{widetext}

\begin{equation}
    \mathcal{N}^{\pm}_{\bm{k}n} = 
    \begin{bmatrix}
        E_{\bm{k}A} \pm i\omega_n + f_{\bm{k}A} & U_{AB} & 0 & 0 \\
        U_{AB} & E_{\bm{k}B} \pm i\omega_n + f_{\bm{k}B} & 0 & 0 \\
        0 & 0 & E_{\bm{k}A} \pm i\omega_n - f_{\bm{k}A} & U_{AB} \\
        0 & 0 & U_{AB} & E_{\bm{k}B} \pm i\omega_n - f_{\bm{k}B} \\
    \end{bmatrix},
\end{equation}
\end{widetext}
\begin{equation}
    \mathcal{N}_{\bm{k}n} = 
    \begin{bmatrix}
        0 & 0 & F_{A} & U_{AB} \\
        0 & 0 & U_{AB} & F_{B} \\
        F_{A} & U_{AB} & 0 & 0 \\
        U_{AB} & F_{B} & 0 & 0 \\
    \end{bmatrix}.
\end{equation}

This yields the partition function 
\begin{equation}
    \mathcal{Z} = e^{-\beta S_0}\int\sideset{}{'}\prod_{\bm{k}\neq 0,n}\mathcal{D}\big[ \Phi_{\bm{k}n} \big] e^{-\Phi^{\dagger}_{\bm{k}n} \mathcal{M}_{\bm{k}n} \Phi_{\bm{k}n}},
\end{equation}
which is a product of Gaussian integrals. The primed product indicates that only half the $\bm{k}$-space and $n>0$ are considered. Performing the integrals gives
\begin{equation}
    \mathcal{Z} = e^{-\beta S_0}\sideset{}{'}\prod_{\bm{k}\neq 0,n}\big[ \textrm{Det}\mathcal{M}_{\bm{k}n} \big]^{-1} = e^{-\beta S_0}\sideset{}{'}\prod_{\bm{k}\neq 0,n}e^{-\textrm{Tr}\ln \mathcal{M}_{\bm{k}n}}.
\end{equation}

To find the contribution from the terms $f_{\bm{k}\alpha}$, the matrix is separated into two pieces $\mathcal{M}_{\bm{k}n} = A_{\bm{k}n} + B_{\bm{k}}$, where $B_{\bm{k}}$ is the diagonal matrix with $f_{\bm{k}\alpha}$ as elements, and $A_{\bm{k}n}$ the remainder of $\mathcal{M}_{\bm{k}n}$. The partition function is then expanded in powers of $B_{\bm{k}}$,
\begin{equation}
    \begin{split}
        \text{Tr}\ln \mathcal{M}_{\bm{k}n} & = \text{Tr}\ln (A_{\bm{k}n} + B_{\bm{k}}) = \text{Tr}\ln A_{\bm{k}n}(I + A^{-1}_{\bm{k}n}B_{\bm{k}}) \\
        &= \text{Tr}\ln A_{\bm{k}n} + \text{Tr}\ln (I + A^{-1}_{\bm{k}n}B_{\bm{k}}) \\
        & \approx \text{Tr}\ln A_{\bm{k}n} + \text{Tr} A^{-1}_{\bm{k}n}B_{\bm{k}} - \frac{1}{2}\text{Tr}(A^{-1}_{\bm{k}n}B_{\bm{k}})^{2},
    \end{split}
\end{equation}
where the power series representation of the natural logarithm, $\ln(1+x) = \sum^{\infty}_{n=1}(-1)^{n+1}\frac{x^{n}}{n}$, has been used to second order. The partition function becomes
\begin{equation}
    \begin{split}
        \mathcal{Z} &= \underbrace{e^{-\beta S_0}\sideset{}{'}\prod_{\bm{k}\neq 0,n} e^{-\text{Tr}\ln A_{\bm{k}n} }}_\text{(i)} \underbrace{ \vphantom{\sideset{}{'}\prod_{\bm{k}}\prod_{n}}  e^{-\text{Tr} A^{-1}_{\bm{k}n}B_{\bm{k}} + \frac{1}{2}\text{Tr}(A^{-1}_{\bm{k}n}B_{\bm{k}})^{2}}}_\text{(ii)} \\
        &= \mathcal{Z}_1 \mathcal{Z}_2.
    \end{split}
\label{eq:Z_PI_split}
\end{equation}
Part (i), $\mathcal{Z}_1$, is the expression for the partition function with $f_{\bm{k}\alpha}=0$ and is solved by a general Bogoliubov transformation (see appendix \ref{app:Diagonalization}), while part (ii), $\mathcal{Z}_2$, is the correction factor due to $f_{\bm{k}\alpha}$ to second order. The first trace of $\mathcal{Z}_2$ vanishes, whereas the second trace is given by
\begin{equation}
    \begin{split}
        \text{Tr}(A^{-1}_{\bm{k}n}B_{\bm{k}})^{2} = & \frac{ 64 f_{\bm{k}A}f_{\bm{k}B}\epsilon_{\bm{k}A}\epsilon_{\bm{k}B}U^2_{AB} (i\omega_n)^2  (1-\frac{1}{2}\delta_{n,0})} {\big[(i\omega_n)^2-\mathcal{E}^2_{\bm{k}+} \big]^2 
        \big[ (i\omega_n)^2 - \mathcal{E}^2_{\bm{k}-} \big]^2 } \\
        &+ \mathcal{O}(f^2_{\bm{k}\alpha}).
    \end{split}
\end{equation}
Moving the product over $n$ inside the exponential, a Matsubara sum of the frequencies $\omega_{n}=2\pi n/\beta$ for $n=\{0,1,2,...\}$ is obtained. Writing out the partial fraction gives
\begin{widetext}

\begin{equation}
	\begin{split}
		\frac{(i\omega_n)^2} {\big[(i\omega_n)^2-\mathcal{E}^2_{\bm{k}+} \big]^2 \big[ (i\omega_n)^2 - \mathcal{E}^2_{\bm{k}-} \big]^2 } = & \frac{\mathcal{E}^2_{\bm{k}+}}{(\mathcal{E}^2_{\bm{k}+} - \mathcal{E}^2_{\bm{k}-})^2 (\omega^2_n + \mathcal{E}^2_{\bm{k}+})^2}
		+ \frac{\mathcal{E}^2_{\bm{k}+} + \mathcal{E}^2_{\bm{k}-}}{(\mathcal{E}^2_{\bm{k}+} - \mathcal{E}^2_{\bm{k}-})^3 (\omega^2_n + \mathcal{E}^2_{\bm{k}+})} \\
		& - \frac{\mathcal{E}^2_{\bm{k}+} + \mathcal{E}^2_{\bm{k}-}}{(\mathcal{E}^2_{\bm{k}+} - \mathcal{E}^2_{\bm{k}-})^3 (\omega^2_n + \mathcal{E}^2_{\bm{k}-})}
		 + \frac{\mathcal{E}^2_{\bm{k}-}}{(\mathcal{E}^2_{\bm{k}+} - \mathcal{E}^2_{\bm{k}-})^2 (\omega^2_n + \mathcal{E}^2_{\bm{k}-})^2},
    \end{split}
\end{equation} 
which has two kinds of sums; one over $(n^2 + a^2)^{-1}$ and $(n^2 + a^2)^{-2}$, both of which can be computed using the identity \cite{Arfken2012}
\begin{equation}
	\sum_{n = 1}^{\infty} \frac{1}{n^2 + a^2} = \frac{\pi}{2a}\coth(\pi a) - \frac{1}{2a}.
\label{eq:starting_point_sum}
\end{equation}
The first is found by adding $1/2a$ to \eqref{eq:starting_point_sum};
\begin{equation}
	\sum_{n = 0}^{\infty} \frac{(1-\frac{1}{2}\delta_{n,0})}{n^2 + a^2} = \frac{\pi}{2a}\coth(\pi a).
\label{eq:sum1}
\end{equation}
The second is found by differentiating \eqref{eq:sum1} with respect to $a$ and rewriting;
\begin{equation}
	\sum_{n = 0}^{\infty} \frac{(1-\frac{1}{2}\delta_{n,0})}{(n^2 + a^2)^2} = \frac{\pi}{4a^3}\coth(\pi a) + \frac{\pi^2}{4a^2}\frac{1}{\sinh^2(\pi a)}.
\end{equation}
The Matsubara sum is therefore

\begin{equation}
	\begin{split}
		\sum^{\infty}_{n = 0} \frac{(1-\frac{1}{2}\delta_{n,0})(i\omega_n)^2} {\big[(i\omega_n)^2-\mathcal{E}^2_{\bm{k}+} \big]^2 
		\big[ (i\omega_n)^2 - \mathcal{E}^2_{\bm{k}-} \big]^2 } = &
		\frac{\beta}{8 (\mathcal{E}^2_{\bm{k}+} - \mathcal{E}^2_{\bm{k}-})^2} \Bigg\{ \frac{\coth(\beta \mathcal{E}_{\bm{k}+}/2)}{\mathcal{E}_{\bm{k}+}} + \frac{\coth(\beta \mathcal{E}_{\bm{k}-}/2)}{\mathcal{E}_{\bm{k}-}} \\
		& + \frac{\beta}{2\sinh^2(\beta\mathcal{E}_{\bm{k}+}/2)} +  \frac{\beta}{2\sinh^2(\beta\mathcal{E}_{\bm{k}-}/2)} \\
		& + 2\frac{(\mathcal{E}^2_{\bm{k}+} + \mathcal{E}^2_{\bm{k}-})}{(\mathcal{E}^2_{\bm{k}+} - \mathcal{E}^2_{\bm{k}-})} \Big[ \frac{\coth(\beta\mathcal{E}_{\bm{k}+}/2)}{\mathcal{E}_{\bm{k}+}} -  \frac{\coth(\beta\mathcal{E}_{\bm{k}-}/2)}{\mathcal{E}_{\bm{k}-}} \Big]  \Bigg\}.
	\end{split}
\label{eq:matsubara_sum_solved}
\end{equation}

In the zero temperature limit, $\beta\rightarrow\infty$, the contribution of $\mathcal{Z}_2$ to the free energy density reduces to
\begin{equation}
	-\frac{1}{\beta N_s}\ln\mathcal{Z}_2 \rightarrow \frac{1}{N_s}\sum_{\bm{k}\neq 0} \frac{2f_{\bm{k}A}f_{\bm{k}B} U^2_{AB}\epsilon_{\bm{k}A} \epsilon_{\bm{k}B}}{\mathcal{E}_{\bm{k}+}\mathcal{E}_{\bm{k}-}(\mathcal{E}_{\bm{k}+} + \mathcal{E}_{\bm{k}-})^3} + \mathcal{O}(f^2_{\bm{k}\alpha}).
\end{equation}

\end{widetext}

\section{Diagonalization of Hamiltonian}
\label{app:Diagonalization}
To diagonalize a general bilinear bosonic Hamiltonian the method of Refs \cite{Xiao2009, vanHemmen1980} is followed, which is a general Bogoliubov transformation.

The $-\bm{k}$ terms are made explicit in \eqref{eq:mean_field_H2_new} and the boson commutation relations used to write it on the form
\begin{equation}
    \widetilde{H}_2 = \frac{1}{4}\sum_{\bm{k}\neq 0}\Phi^{\dagger}_{\bm{k}}\mathcal{A}_{\bm{k}}\Phi_{\bm{k}} - \frac{1}{2}\sum_{\bm{k}\neq 0}\sum_{\alpha}E_{\bm{k}\alpha},
\end{equation}
where the basis vector $\Phi_{\bm{k}}$ is in the two-component case
\begin{equation}
    \begin{split}
        \Phi^{\text{2-comp.}}_{\bm{k}} = & \Big( b_{\bm{k}A} ,\, b_{\bm{k}B} ,\, b^{\dagger}_{-\bm{k}A} ,\, b^{\dagger}_{-\bm{k}B} ,\, \\
        & b^{\dagger}_{\bm{k}A} ,\, b^{\dagger}_{\bm{k}B},\,  b_{-\bm{k}A} ,\, b_{-\bm{k}B}\Big)^{\textrm{T}},
    \end{split}
\end{equation}
and in the three-component case
\begin{equation}
    \begin{split}
        \Phi^{\text{3-comp.}}_{\bm{k}} = & \Big( b_{\bm{k}A} ,\, b_{\bm{k}B} ,\, b_{\bm{k}C} ,\, b^{\dagger}_{-\bm{k}A} ,\, b^{\dagger}_{-\bm{k}B} ,\, b^{\dagger}_{-\bm{k}C} ,\, \\
        & b^{\dagger}_{\bm{k}A} ,\, b^{\dagger}_{\bm{k}B} ,\, b^{\dagger}_{\bm{k}C} ,\,  b_{-\bm{k}A} ,\, b_{-\bm{k}B} ,\, b_{-\bm{k}C} \Big)^{\textrm{T}}.
    \end{split}
\end{equation}
Both of the above yields the matrix $\mathcal{A}_{\bm{k}}$ on the form
\begin{equation}
    \mathcal{A}_{\bm{k}} = 
    \begin{bmatrix}
        \mathcal{N}_{\bm{k}} & 0 \\
        0 & \mathcal{N}_{\bm{k}}
  \end{bmatrix},
\end{equation}
with 
\begin{widetext}
\begin{equation}
    \mathcal{N}^{\text{2-comp.}}_{\bm{k}} =
    \begin{bmatrix}
        E_{\bm{k}A}+f_{\bm{k}A} & U_{AB} & F_{A} & U_{AB} \\
        U_{AB} & E_{\bm{k}B}+f_{\bm{k}B} & U_{AB} & F_{B} \\
        F_{A} & U_{AB} & E_{\bm{k}A}-f_{\bm{k}A} & U_{AB} \\
        U_{AB} & F_{B}& U_{AB} & E_{\bm{k}B}-f_{\bm{k}B}\\
  \end{bmatrix},
\end{equation}
for two-components, and
\begin{equation}
    \mathcal{N}^{\text{3-comp.}}_{\bm{k}} =
    \begin{bmatrix}
        E_{\bm{k}A}+f_{\bm{k}A} & U_{AB} & U_{AC} & F_{A} & U_{AB} & U_{AC} \\
        U_{AB} & E_{\bm{k}B}+f_{\bm{k}B} & U_{BC} & U_{AB} & F_{B} & U_{BC} \\
        U_{AC} & U_{BC} & E_{\bm{k}C}+f_{\bm{k}C} & U_{AC} & U_{BC} & F_{C} \\
        F_{A} & U_{AB} & U_{AC} & E_{\bm{k}A}-f_{\bm{k}A} & U_{AB} & U_{AC} \\
        U_{AB} & F_{B} & U_{BC} & U_{AB} & E_{\bm{k}B}-f_{\bm{k}B} & U_{BC} \\
        U_{AC} & U_{BC} & F_{C} & U_{AC} & U_{BC} & E_{\bm{k}C}-f_{\bm{k}C} \\
  \end{bmatrix}.
\end{equation}
\end{widetext}
for three-components.
The transformation into the diagonal basis $c_{\bm{k}i}$ must preserve the boson commutation relations, which in terms of $\Phi_{\bm{k}}$ reads
\begin{equation}
    \left[ \Phi, \Phi^{\dagger} \right] = \Phi \Phi^{\dagger} - \left( \Phi^{\dagger} \Phi \right)^{\textrm{T}} = \mathcal{J} = \text{diag}\big(\widetilde{\mathcal{J}}, \, -\widetilde{\mathcal{J}}\big),
\end{equation}
where $\widetilde{\mathcal{J}} = \text{diag}(1, 1, -1, -1)$
\cite{Tsallis1978}. Thus, to find the bosonic energy spectrum the matrix $\mathcal{A}_{\bm{k}}\mathcal{J}$ is diagonalized to yield $\mathcal{D}_{\bm{k}}\mathcal{J}$, where $\mathcal{D}_{\bm{k}} = \text{diag}(\mathcal{E}_{\bm{k}1},... \mathcal{E}_{\bm{k}n}, \mathcal{E}_{\bm{k}1},...,\mathcal{E}_{\bm{k}n})$ with the energies $\mathcal{E}_{\bm{k}i}$ and $n$ either 4 or 6 for two or three components, respectively. Note that the diagonalization yields the energies of $\bm{k}$ and $-\bm{k}$ simultaneously, e.g. $\mathcal{E}_{\bm{k}1} = \mathcal{E}_{-\bm{k}3}$. $\mathcal{E}_{\bm{k}2} = \mathcal{E}_{-\bm{k}4}$, $c_{\bm{k}1}=c_{-\bm{k}3}$, and $c_{\bm{k}2}=c_{-\bm{k}4}$ in the two-component case.
Furthermore, since $\mathcal{A}_{\bm{k}}$ is in block diagonal form only the diagonalization of $\mathcal{N}_{\bm{k}}\widetilde{\mathcal{J}}$ needs to be considered.

This yields \eqref{eq:mean_field_H2} in the diagonal basis
\begin{equation}
    \begin{split}
        \widetilde{H}_2 = & \sum_{\bm{k}\neq 0}\sum_{i}\mathcal{E}_{\bm{k}i}c^{\dagger}_{\bm{k}i}c_{\bm{k}i}
        + \frac{1}{2}\sum_{\bm{k}\neq 0}\sum_{i}\mathcal{E}_{\bm{k}i} \\
        & - \frac{1}{2}\sum_{\bm{k}\neq 0}\sum_{\alpha}E_{\bm{k}\alpha}.
    \end{split}
\end{equation}

\section{Rayleigh-Schrödinger Perturbation Theory}
\label{app:RS_perturbation_theory}
In Rayleigh-Schrödinger perturbation theory, when the Hamiltonian has been separated into an exactly solvable part $H_{\text{sol}}$ and a perturbation $H_{\text{pert}}$, the system energy and the state are expanded in a smallness parameter $\lambda$,
\begin{equation}
    E = E^{\text{(0)}} + E^{\text{(1)}} + E^{\text{(2)}} + \dots,
\end{equation}
\begin{equation}
    \ket{\Psi} = \ket{\Psi^{\text{(0)}}} + \ket{\Psi^{\text{(1)}}} + \ket{\Psi^{\text{(2)}}} + \dots,
\end{equation}
where the terms $E^{\text{(i)}}$ and $\ket{\Psi^{\text{(i)}}}$ are of order $\lambda^i$.
The zeroth order states and energies are those of the exactly solvable system, while the higher orders are corrections. To find the expressions for the corrections. the Schrödinger equation is solved recursively in a standard manner using the above expansion, see for instance \cite{Griffiths2005}. In order to compute the drag-coefficients, obtaining the corrections to the energy will suffice. To first, second, third, and fourth order we find
\begin{equation}
    E^{\text{(1)}} = V_{00},
\label{eq:first_order_RS}
\end{equation}
\begin{equation}
    E^{\text{(2)}} = \sum_{m\neq 0} \frac{|V_{0m}|^2}{E_{0m}},
\label{eq:second_order_RS}
\end{equation}
\begin{equation}
    E^{\text{(3)}} = \sum_{ml\neq 0} \frac{V_{0m}V_{ml}V_{l0}}{E_{0m}E_{0l}} - V_{00}\sum_{m\neq 0}\frac{|V_{0m}|^2}{E_{0m}^2},
\label{eq:third_order_RS}
\end{equation}
\begin{equation}
    \begin{split}
        E^{\text{(4)}} = &\sum_{mlr\neq 0} \frac{V_{0m}V_{ml}V_{lr}V_{r0}}{E_{0m}E_{0l}E_{0r}} - \sum_{ml\neq 0}\frac{|V_{0m}|^2}{E_{0m}}\frac{|V_{0l}|^2}{E_{0l}^2}\\
        &- 2V_{00}\sum_{ml\neq0}\frac{V_{0m}V_{ml}V_{l0}}{E_{0m}E_{0l}^2} + V_{00}\sum_{m\neq 0} \frac{|V_{0m}|^2}{E_{0m}^3},
    \end{split}
\label{eq:fourth_order_RS}
\end{equation}
using the notation 
\begin{equation}
    \begin{split}
        &V_{ml} = \bra{N_m} H_{\text{pert}} \ket{N_l}, \\
        &E_{ml} = E_m - E_l,
    \end{split}
\end{equation}
where $\ket{N_m}$ and $\ket{N_l}$ are states in Fock space with corresponding energies $E_m$ and $E_l$. The $0$ index indicates the unperturbed state $\ket{\Psi^{\text{(0)}}}$. When this is equal to the ground state $\ket{GS}$, as in this paper, the energy difference $E^{\text{(0)}} - E_m$ will always be negative.

A criterion for when this perturbative expansion is useful is for the correction terms to be small compared to the energy levels,
\begin{equation}
    |V_{ml}| \ll |E_{ml}|.
\end{equation}

\section{Details of the Numerics}
The algorithm used in the simulations is a multi-component version of the worm algorithm outlined by Ref \cite{PingNangMa} based on Ref \cite{Pollet2007}. Starting from an empty configuration, $10^6-10^7$ worm insertions were conducted in order to thermalize the system. A large number of samples was found to be necessary in order to obtain reliable statistics for the winding numbers. The data points were therefore produced using $10^8-10^9$ samples with $10^1-10^2$ worm insertions between each sample. The normalization data point was obtained using $\mu_A = -3.6053$, $\mu_B = -3.6068$, $\mu_C = -3.7468$ producing $n_A = 0.301$, $n_B = 0.297$ and $n_C = 0.300$. The rest of the chemical potentials used are displayed in table \ref{Table1} and \ref{Table2}. Starting from values provided by mean-field theory, the values for the chemical potentials were determined through iterative simulations until a satisfactory accuracy was achieved. \\
\vspace{0.5cm}

\begin{table}[H]
\setlength{\tabcolsep}{4.5pt}
\def\arraystretch{1.5}
\begin{tabular}{ c  c  c  c  c  c  c }
\hline
\hline
$\lambda_{AC}$ & $\mu_A$ & $\mu_B$ & $\mu_C$ & $n_A$ & $n_B$ & $n_C$\\
\hline
-0.25 & -3.6797 & -3.4594 & -3.6797 & 0.303 & 0.301 & 0.303\\
0     & -3.6024 & -3.4595 & -3.6021 & 0.301 & 0.302 & 0.302\\
0.25  & -3.5303 & -3.4609 & -3.5294 & 0.300 & 0.300 & 0.301\\
0.5   & -3.4591 & -3.4596 & -3.4590 & 0.301 & 0.300 & 0.302\\
\hline
\hline
\end{tabular}
\caption{Values for the chemical potentials and the resulting particle densities for each of the data points in \figref{fig:Drag_MF_VS_QMC} with $\lambda_{BC}=0.5$.}
\label{Table1}
\end{table}

\begin{table}[H]
\setlength{\tabcolsep}{4.5pt}
\def\arraystretch{1.5}
\begin{tabular}{ c  c  c  c  c  c  c }
\hline
\hline
$\lambda_{AC}$ & $\mu_A$ & $\mu_B$ & $\mu_C$ & $n_A$ & $n_B$ & $n_C$\\
\hline
-0.5  & -3.7609 & -3.7604 & -4.0638 & 0.299 & 0.300 & 0.298\\
-0.25 & -3.6802 & -3.7621 & -3.9811 & 0.301 & 0.300 & 0.300\\
0     & -3.6018 & -3.7599 & -3.9068 & 0.302 & 0.301 & 0.299\\
0.25  & -3.5301 & -3.7593 & -3 8348.& 0.300 & 0.304 & 0.300\\
\hline
\hline
\end{tabular}
\caption{Values for the chemical potentials and the resulting particle densities for each of the data points in \figref{fig:Drag_MF_VS_QMC} with $\lambda_{BC}=-0.5$.}
\label{Table2}
\end{table}

\bibliographystyle{apsrev4-1}  
\addcontentsline{toc}{chapter}{\bibname}
\bibliography{Refs}  

\end{document}